\documentclass{emulateapj}

\slugcomment{Submitted: 2006/05/10}
\slugcomment{Accepted: July 20, 2006}

\shorttitle{IR Observations of the Helix}
\shortauthors{Hora et al.}

\begin{document}


\title{Infrared Observations of The Helix Planetary Nebula
    }


\author{Joseph L. Hora\altaffilmark{1}, William B. Latter\altaffilmark{2},
 Howard A. Smith\altaffilmark{1},
Massimo Marengo\altaffilmark{1}
}

\altaffiltext{1}{Harvard-Smithsonian Center for Astrophysics, 60 Garden St., 
MS-65, Cambridge, MA  02138-1516}
\altaffiltext{2}{NASA/Herschel Science Center, MS 220-6, 
California Institute of 
Technology, Pasadena, CA 91125}
\email{jhora@cfa.harvard.edu}



\begin{abstract} 

We have mapped the Helix (NGC 7293) planetary nebula (PN) with the IRAC
instrument on the {\it Spitzer Space Telescope}.  The Helix is one of the
closest bright PN, and therefore provides an opportunity to resolve the
small-scale structure in the nebula.  The emission from this PN in the 5.8
and 8 micron IRAC bands is dominated by the pure rotational lines of
molecular hydrogen, with a smaller contribution from forbidden line
emission such as [\ion{Ar}{3}] in the ionized region.  The IRAC images
resolve the ``cometary knots'' which have been previously studied in this
PN.  The ``tails'' of the knots and the radial rays extending into the
outer regions of the PN are seen in emission in the IRAC bands.  IRS
spectra on the main ring and the emission in the IRAC bands are consistent
with shock-excited H$_2$ models, with a small ($\sim$10\%) component from
photodissociation regions. In the Northeast Arc, the H$_2$ emission is
located in a shell outside of the H$\alpha$ emission.

\end{abstract}





\section{Introduction}

One of the remarkable features of The Helix (NGC 7293; PN G036.1-57.1)
planetary nebula (PN) is the small-scale structure in its expanding
shells.  The appearance is more striking because of the close proximity
\citep[213 pc;][]{harris97} of the Helix compared to other PN, giving us a
clearer view of the nebula's structure.  The distribution of material in
the Helix is unlike the simple classical picture of a uniform spherical
shell expanding isotropically from the central star.  Instead, the
material in the shells is seen to reside in cometary knots and clumps, and
the structure of these clumps changes as a function of distance from the
central star.  \citet{odell04} combined {\it Hubble Space Telescope} (HST)
and ground-based imaging to provide a comprehensive picture of the Helix.  
Closest to the central star are the cometary knots, which have highly
ionized emission along their rims nearest the central star and fainter
emission along the edges of the material that is trailing away from the
star.  Further from the central star are the main rings of the nebula, in
which the knots are closely spaced and not as sharply defined as the inner
cometary knots.  Outside of the main rings, faint rays of emission extend
out to a diffuse outermost ring at a radius approximately 15 arcmin from
the central star.  \citet{odell04} conclude that the nebular
structure consists of an inner disk roughly in the plane of the sky,
surrounded by a highly inclined torus, with an outer ring at roughly twice
the diameter of the inner structures.

The orientation of the Helix provides an excellent view of the region
around the central star that has been cleared of molecular material and
allows us to view the cometary knots nearly in profile.  The cometary
knots have been studied extensively, more recently with HST imaging and
spectra providing the best resolution of the structures
\citep[e.g.,][]{meaburn92,odell96,odell97, odell05,meix05}.  The knots
typically have a bright cusp of emission in lines H$\alpha$ and forbidden
lines such as [\ion{O}{1}] and [\ion{N}{2}].  In [\ion{O}{3}] the knots
and their tails are visible as shadows in the extended emission around the
knot.  In the 2.12 $\mu$m H$_2$ line, the emission is also brightest along
the rim facing the central star \citep{huggins92,meix05}, but there is
significant emission from the tail in H$_2$ and CO. The structure and
kinematics of the knots, and their change in morphology from the inner to
outer regions is consistent with the knots forming near the ionization
front and being shaped by the stellar radiation as the ionization front
moves outward \citep{odell04}.  However, models of the H$_2$ emission from
the knots \citep{odell05, meix05} fail to accurately reproduce the flux
observed in the near-IR lines.

The ISOCAM instrument on board the {\it Infrared Space Observatory} ({\it
ISO}) was used to produced infrared images and spectra of the Helix in the
5 -- 17 $\mu$m spectral region \citep{cox98}.  The nebula was mapped with
6\arcsec\ resolution and mJy sensitivity, and they found that the emission
was dominated by the pure rotational lines of H$_2$ in the 5 -- 12 $\mu$m
region, and by the 12.81 $\mu$m [\ion{Ne}{2}] and 15.55 $\mu$m
[\ion{Ne}{3}] line at longer wavelengths.  Conspicuously absent was any
emission from polycyclic aromatic hydrocarbons (PAHs) which one might
expect from what they assumed was a carbon-rich nebula. 

In this paper, we report results based on observations of the Helix PN
using the Infrared Array Camera \citep[IRAC,][]{fazio04} and the Infrared
Spectrograph \citep[IRS,][]{houck04} on board the {\it Spitzer Space
Telescope} \citep{werner04}.  With its higher resolution ($\sim$2\arcsec)
and sensitivity (5$\sigma$ extended source sensitivity at 8 $\mu$m of 0.2
MJy sr$^{-1}$ in 180 sec), and wider field (5.2\arcmin$\times$5.2\arcmin)
than previous cameras in this wavelength range, IRAC is a powerful
instrument for investigating nebular emission from gas and dust in PN,
reflection nebulae, and star-forming regions.  The four IRAC channels
sample the wavelength range from $\sim$ 3.1 -- 9.5 $\mu$m, which
potentially includes emission lines from ionized gas such as Br$\alpha$ at
4.05 $\mu$m, forbidden line emission such as the [\ion{Mg}{5}] 5.6 $\mu$m,
[\ion{Ne}{6}] 7.64 $\mu$m, and [\ion{Ar}{3}] 8.99 $\mu$m lines, emission
from H$_2$ from transitions in all four bands, CO emission from
transitions near 4.65 $\mu$m, broad features such as the PAH lines at 3.3,
6.2, 7.7, and 8.6 $\mu$m, and continuum emission from warm or hot dust.  
Before {\it ISO} and {\it Spitzer}, previous observations of PN in this
wavelength range that were obtained from the ground were primarily of
young objects, e.g., NGC 7027 \citep{aitken83} and BD+30$^{\circ}$3639
\citep{hora93}, that were strong mid-IR sources due to their PAH or warm
dust emission. With IRAC, the mid-IR emission from more evolved PN can be
investigated.  We are conducting a program to observe a sample of 35 PN
with IRAC.  Initial results were reported in \citet{hora04,hora05}, and
showed IRAC was especially sensitive to the ionized gas emission in the
nebulae and could detect faint H$_2$ emission in the outer shells and
halos.  In this paper we show the results for the Helix, which is the PN
with the largest angular extent in the sample.

\section{Observations and Data Reduction}


\subsection{IRAC images}

The observations were obtained with the IRAC instrument \citep{fazio04} on
the {\it Spitzer Space Telescope} \citep{werner04}.  The 30-sec ``High
Dynamic Range'' (HDR) mode was used (AOR 0004422400).  The HDR mode takes
pairs of images with 1.2 and 30 sec frame times (0.6 and 26.8 sec exposure
time)  at each dither position in the IRAC channels 1 -- 4 (3.6, 4.5, 5.8
and 8 $\mu$m; see \citet{fazio04} for the band transmissions and
isophotal wavelengths).  A 5$\times$5 map was performed with 6 dither
positions per map position, resulting in a median exposure time of
$\sim$160 seconds. The S11.4.0 version of the Basic Calibrated Data (BCD)
products from the Spitzer Science Center (SSC)  pipeline were used to
construct the mosaic images. The BCD products have the main instrumental
signatures removed from the data and are calibrated in units of MJy
sr$^{-1}$, based on the calibration derived from measurements of standard
stars \citep{fazio04, reach05}.  Some bright source residual artifacts in
the pipeline images caused by bright stars were removed by forcing the
column or row median in regions with no sources to be equal to that of
adjacent columns or rows.  Then the individual BCD images were combined
into a single image for each channel and frame time using the SSC
``mopex'' mosaicer program.  The ``IRAC\_proc'' version 3.0 scripts
developed by \citet{schust06} were used to determine the pixel masks and
run the mopex software. The output images were written with a linear pixel
size 1/3 that of the input pixels (1/9 of the area). For flux calibration,
it was assumed that the zero magnitude fluxes in the IRAC bands are 277.5.
179.5, 116.6, and 63.1 Jy for channels 1-4, respectively.  A flux
correction for the extended source emission was applied to the extracted
fluxes used in this paper, as described in the Spitzer Observer's
Manual\footnote{http://ssc.spitzer.caltech.edu/documents/som/} and
\citet{reach05}.  For the extended emission fluxes presented here, the MJy
sr$^{-1}$ values in the images from the pipeline were multiplied by
factors of 0.944, 0.937, 0.772, and 0.737 for IRAC channels 1 -- 4,
respectively. The local sky background in each IRAC band was estimated
from regions outside of the nebula, and subtracted from the image.

\subsection{Ground-based Near-IR Molecular Hydrogen Image} 

A narrowband near-IR image at 2.12 $\mu$m was obtained on 1997 June 18
using the QUIRC camera \citep{hodapp96} and the Quick Infrared Survey
Telescope (QUIST) on Mauna Kea. QUIST is a f/10 Ritchey-Chretien
Cassegrain telescope with a 25.4 cm diameter primary that was attached to
the top of QUIRC, and the system was mounted on the University of Hawaii
61 cm telescope on Mauna Kea for pointing. The observing for this project
was controlled remotely from Kaneohe, HI, with the assistance of the UH
2.2m telescope operator to open and close the dome, and refill the
camera's LN$_2$ supply. The QUIST telescope with QUIRC provided a pixel
scale of 1\farcs68 pixel$^{-1}$ and a field-of-view of approximately
29\arcmin$\times$29\arcmin.  A 5-position dither pattern of 120 sec frames
was used that placed the nebula in the center of each quadrant and the
center of the array, covering a roughly 1 square degree area.  A total of
48 frames were obtained and mosaiced to form the final image.  The QUIRC
narrowband H$_2$ filter is roughly Gaussian in shape, centered at 2.124
$\mu$m with a bandwidth of 0.022 $\mu$m.

\subsection{IRS Spectra}

The spectra are from a calibration dataset (AOR 0013736192) obtained on
2005 May 29 with the IRS. The low-resolution spectra used a ramp duration
of 60 seconds with two cycles.  Two positions were obtained on the main
ring of the nebula, and one position roughly 6\arcmin\ north of the ring.  
The S12.0.2 version of the BCD were used in the reduction.  The data at
each of the positions were averaged separately. Then the northern position
spectral image was subtracted from the ring positions before they were
extracted to remove the background which is dominated by the zodiacal
emission. The SPICE software (version v1.1-beta16) from the SSC was used
to extract the spectra, using the full slit width. The IRS calibration is
based on observations of point sources, so a slit loss correction factor
was applied to normalize the spectra \citep{kelley05}.  This correction
should provide for accurate line ratios, however it may slightly affect
the absolute calibration of the spectrum. The line fluxes were measured
using the spectral analyis routines in the version 5.5 SMART IDL package
written by the IRS team, which can be downloaded from the SSC web site.

\section{Results}
\subsection{IRAC Images}

\begin{figure}
\includegraphics[scale=1.1, angle=-90]{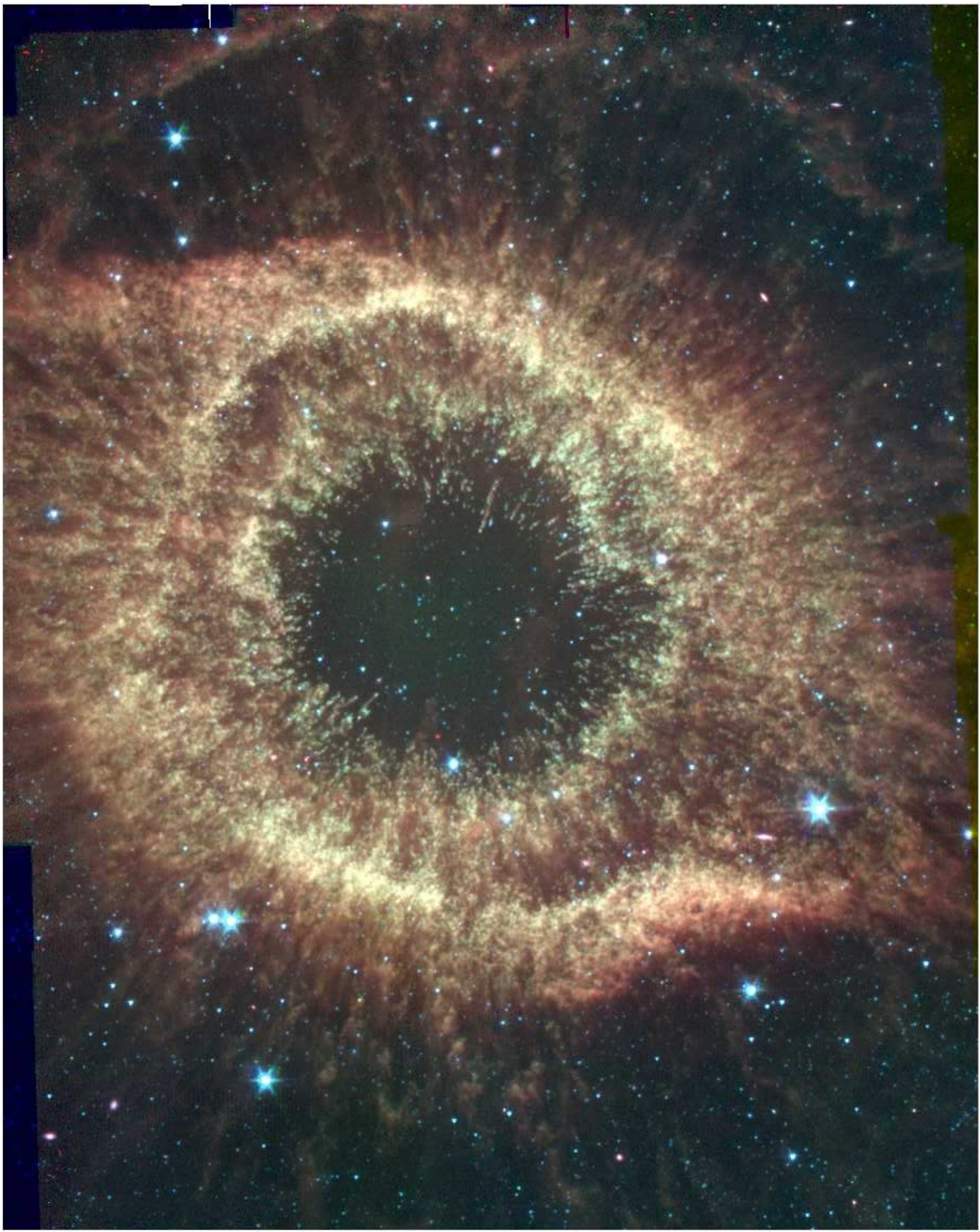}
\caption{Three-color image of the Helix, showing the regions that were
imaged in all IRAC bands.  The 3.6, 4.5, and 8 $\mu$m IRAC
bands are mapped to blue, green, and red, respectively.  The
orientation of the mosaic is in array coordinates, the vertical axis is
59.6 degrees East of North (see Figure \ref{IRAC_all}). The image is
approximately 24.2 $\times$ 26.2 arcmin in size.\label{4colfull} }
\end{figure}

\begin{figure}
\includegraphics[scale=1.1, angle=-90]{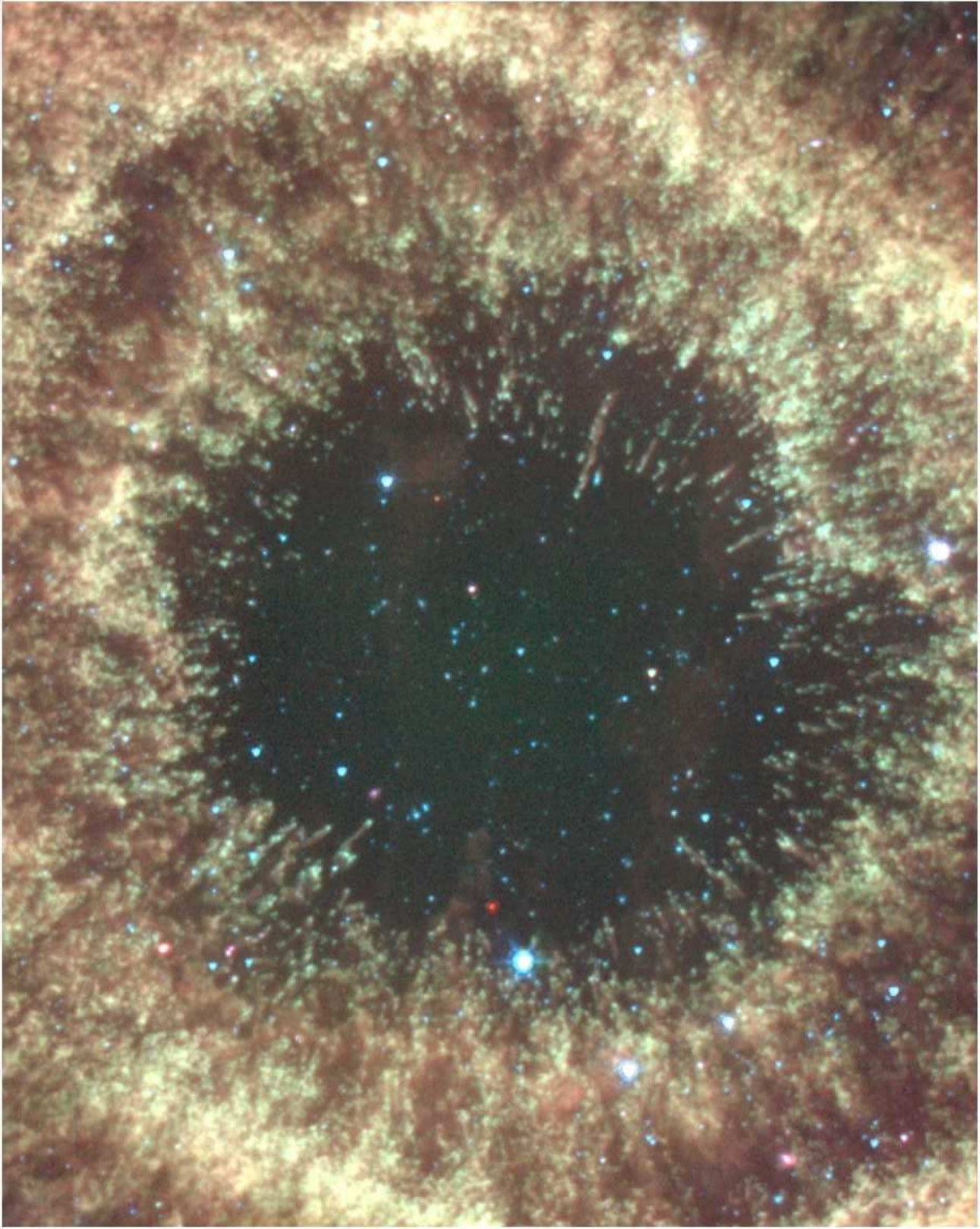}
\caption{
Same as in Fig. 1, except showing the 12 $\times$ 11 arcmin region around
the central star.  This shows the structure of the knots inside the main 
nebular ring.\label{4colfullz}
}
\end{figure}

\begin{figure}
\includegraphics[scale=0.9, angle=0]{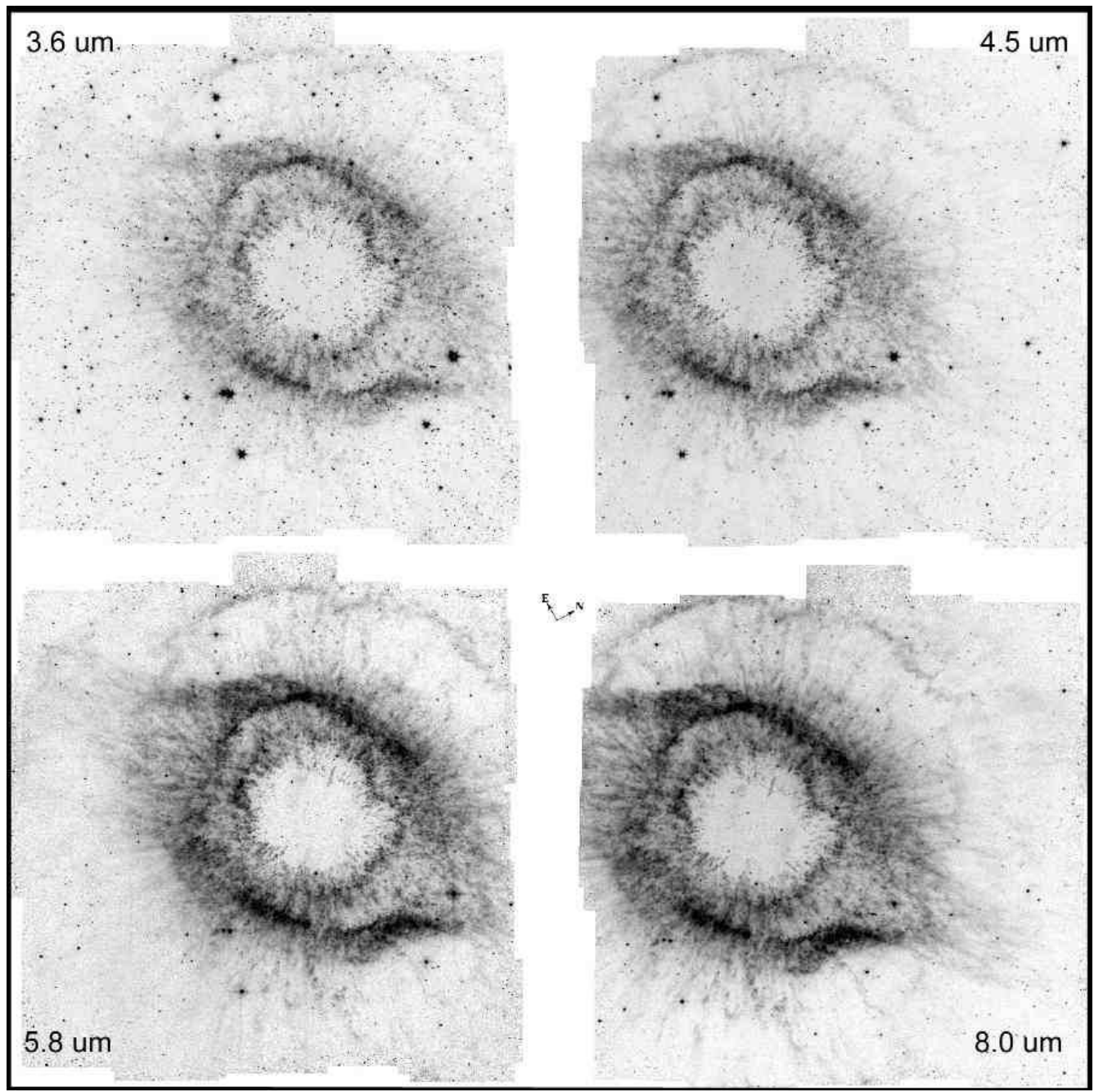}
\caption{
Grayscale images of each of the individual IRAC bands.  The images are labeled
with the name of the IRAC bands,
and the orienation indicated by the arrows in the center of the image.  The 
length of the arrows is 1 arcmin.  The orientation of all the bands is the 
same, and each image is aligned vertically and horizontally with its neighbors.
\label{IRAC_all}}
\end{figure}

\begin{figure}
\includegraphics[scale=0.9, angle=0]{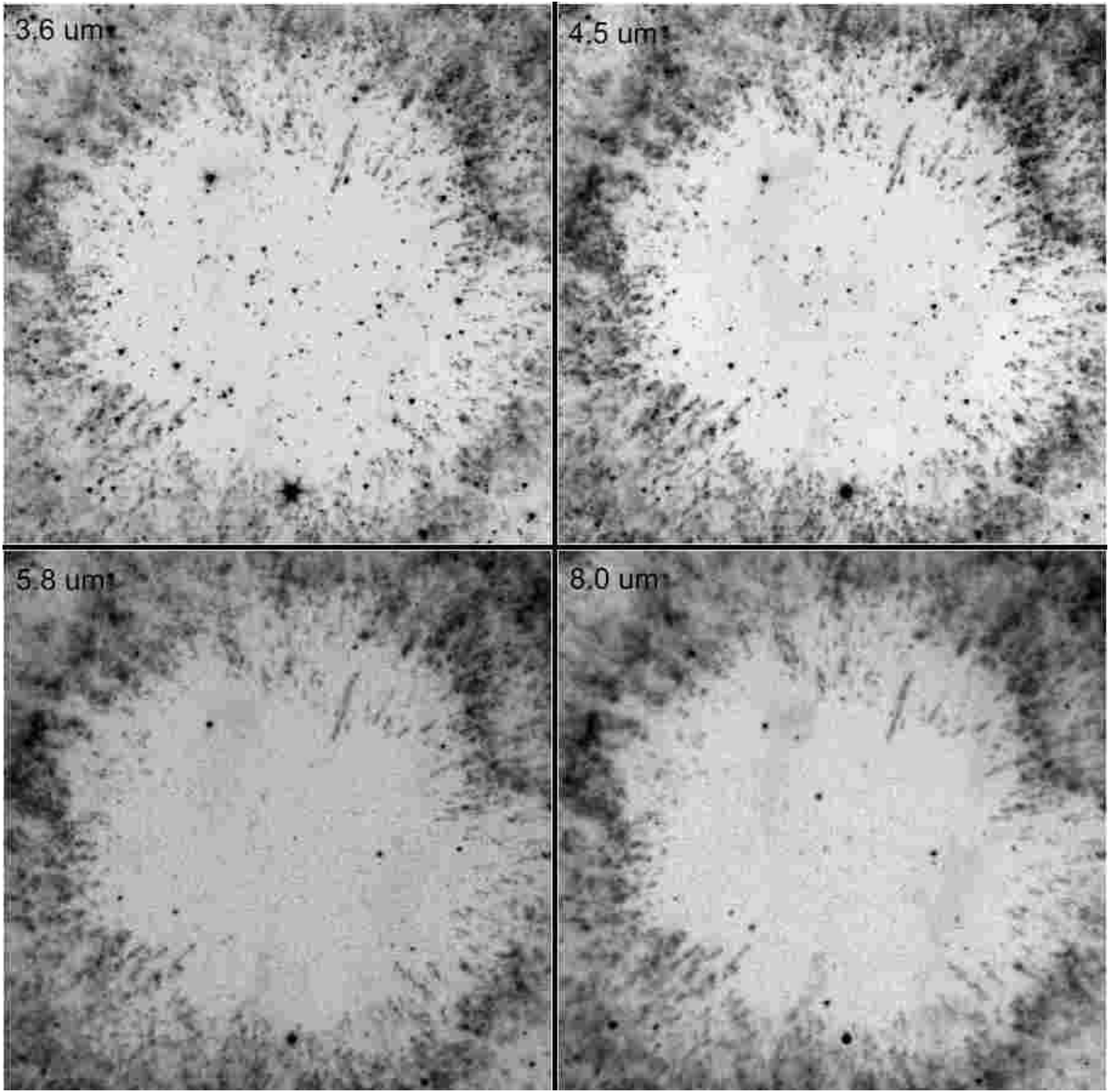}
\caption{
Same as Figure \ref{IRAC_all}, except just the inner 6$\times$6 
arcmin region displayed. \label{IRAC_zoom} 
}
\end{figure}

The IRAC images of the Helix are shown in Figures \ref{4colfull} --
\ref{IRAC_zoom}.  Figure \ref{4colfull} is a color image of three of the IRAC
bands, as described in the figure caption. Figure \ref{4colfullz} shows
the core region to better display the structure of the cometary knots.
Grayscale images of the individual IRAC bands are shown in Figure
\ref{IRAC_all}, and the inner 6\arcmin\ are shown in Figure
\ref{IRAC_zoom}.

One of the most striking features of the color images is the cometary
knots in the inner part of the nebula.  In the optical images of the
cometary knots as shown for example in \citet{odell05}, the knots show a
bright rim or cusp, and a shadowed region appears behind it.  The
brightest part of the knot is the surface that faces the central star.  
The tail is outlined in faint emission, and appears limb-brightened, so
that the outer edges are brighter than the center of the tail. In the IRAC
images, the tips of the knots are brighter in the 3.6 and 4.5 $\mu$m
bands. In the 5.8 and 8 $\mu$m bands, the emission appears relatively
constant along the knot, with little or no brightening at the tip.  This
is apparent in the color images of Figure \ref{4colfull} and
\ref{4colfullz}, where the knots have blue-green tips, and relatively
redder tails.

Another feature of the images is that the emission in the IRAC bands is
fragmented or clumpy throughout the nebula.  The characteristics of the
clumps varies in a systematic way as a function of radius in the nebula.  
In the inner portion there are isolated cometary knots with long tails.  
The heads of the knots are not resolved in the IRAC images.  In the inner
and outer rings, there are also small clumps of emission, but with little
or no tails seen.  Beyond the outer ring, there are rays of emission that
extend to the outermost ring.  Also in this region are faint wisps of
emission many arcsec across that appear like bowshock regions, but much
larger than the cometary knots in the inner regions.  The clumpy structure
in the rings is consistent with that seen by \citet{speck02} and
\citet{meix05} in their H$_2$ images of the Helix.

\subsubsection{IRAC fluxes and colors of the nebula}

Table \ref{fluxtab} lists the fluxes and magnitudes of the main rings of
the nebula.  The halo region was not included since not enough area was
imaged in any of the bands to cover the entire halo.  We restricted the
flux calculation to an elliptical region around the nebula with a major
and minor axes of 1185\arcsec\ and 822\arcsec, with the major axis
oriented 60$\arcdeg$ west of north to align it to the longest dimension of
the main rings.  The bright stars 2MASS J22290943-2046073, 2MASS
J22292575-2056519, and 2MASS J22292663-2057075 ([K] of 7.5 to 9.5) that
are in the outer parts of the ring were masked out from the flux total.
The contribution from field stars was estimated by measuring the total
flux from stars in an area outside of the halo in each of the IRAC bands,
normalizing it to the area of the nebula, and subtracting that from the
total flux measured inside the ellipse.

\begin{table}
\begin{center}
\caption{IRAC Fluxes and Colors of the Main Rings\label{fluxtab}}
\begin{tabular}{crr}
\\
\tableline\tableline
IRAC band & Flux & Magnitude \\
($\mu$m) & (Jy) \\
\tableline
3.6 & 1.71 & 5.54  \\
4.5 & 3.35 & 4.32  \\
5.8 & 6.07 & 3.19  \\
8 & 9.66 & 2.06  \\
\tableline
\end{tabular}
\end{center}
\end{table}

\begin{figure}
\includegraphics[scale=0.5]{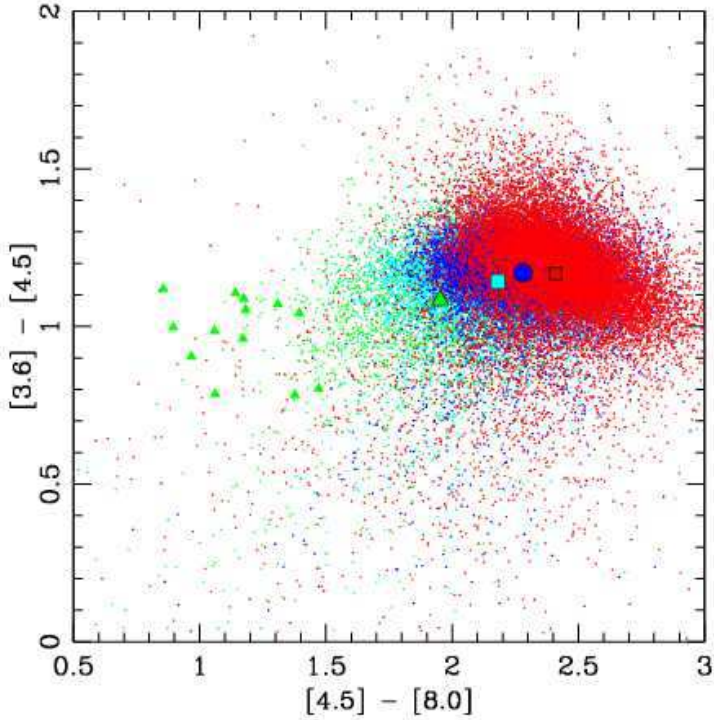}
\caption{
Plotted are the IRAC [3.6] - [4.5] vs. [4.5] - [8] color of the nebula.  
The small dots are the binned (2''$\times$2'') pixels above the minimum flux
cutoff.  The green dots are those in the inner 200'', the cyan dots are 
in the range 220 - 300'', blue are 320 - 400, and red are $>$400'' from the
center of the nebula.  Larger green, cyan, blue, and red symbols show the 
median of the distribution for the radius range.  Smaller green filled 
triangles that appear left of center are the colors of the tips of 
isolated cometary knots, showing that they are much brighter in the 4.5 $\mu$m
band than the rest of the nebula.
\label{n7293_mags}
}
\end{figure}

The IRAC colors of the nebular emission are plotted in Figure
\ref{n7293_mags}.  To make this plot, the IRAC images were binned to
2\arcsec$\times$2\arcsec\ pixels, and only pixels where the 4.5 $\mu$m
band surface brightness was greater than 0.09 MJy sr$^{-1}$ were included
in the calculation.  Also, regions affected by the brightest stars were
masked, and the region interior to the cometary knots was excluded.  The
color of the IRAC emission at each point was then calculated and plotted
in groups of increasing radius bins.  The first bin includes the region
inside the first ring, including the cometary knots. The second bin (cyan)  
contains the first ring, the third bin (blue) contains the second ring,
and the last bin (red) includes the region exterior to the second ring,
out to the limit of the detected emission, but not including the halo or
arcs outside of the outer ring.  No attempt was made to remove the
background stars and galaxies that are visible in the images -- they are
contributing to the scatter of points in the diagram, but will be
relatively small in number and should be uniformly distributed in the
images, so they will not affect the relative colors of the regions in the
nebula.

One can immediately see that the IRAC colors become redder in [4.5] -
[8] as a function of increasing radius.  There is very little change in
the [3.6] - [4.5] color over the same range.  The tips of the cometary
knots are the extreme of the color range, with a median [4.5] - [8]
color of 1.1, compared to the rest of the nebula with medians that range
from $\sim$ 2 to 2.5.  The [5.8] - [8] colors do not show this same
trend.  The regions outside of 100\arcsec\ from the center have the same
color, with a median value of 1.17.  The inner 100\arcsec\ is slightly
redder, with a median value of 1.3.

\subsubsection{Comparison with NICMOS and other data}

\begin{figure}
\includegraphics[scale=0.35]{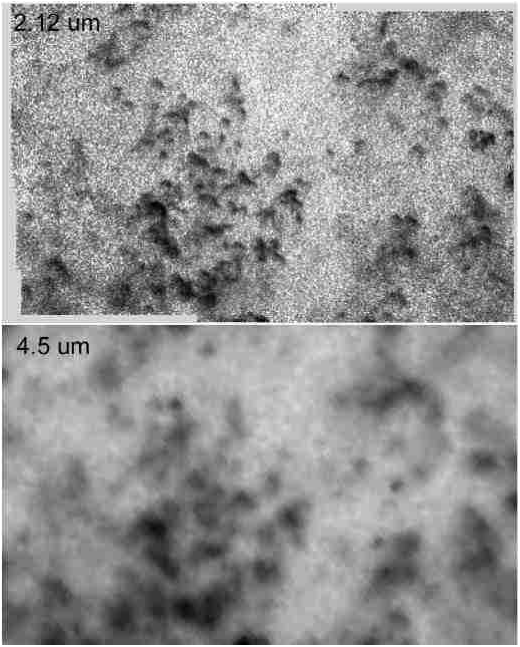}
\caption{
Top: The NICMOS 2.12 $\mu$m H$_2$ image (\citet{meix05}, ``Position 2''). The 
total field is approximately 80$\times$50\arcsec, and N is 21.89 CW from 
vertical.  Bottom:
the same area of the IRAC 4.5 $\mu$m image.
\label{ch2_nicmos}
}
\end{figure}

Figure \ref{ch2_nicmos} shows a comparison between a NICMOS image in the
ring \citep{meix05} and the same region from the IRAC 4.5 $\mu$m image.  
Except for the better resolution of the NICMOS image, the images are very
similar, with each feature in the NICMOS image matching a feature in the
IRAC image, and at approximately the same relative intensity.  There are a
few features in the IRAC image that appear different, and are probably
background stars or galaxies. These are brighter in the IRAC images due to
the broader bandpass of the IRAC filters compared to the narrowband H$_2$
filter used for the NICMOS image, and also the extragalactic sources are
relatively brighter at the longer wavelength of the IRAC image.
 
The close correspondence between the NICMOS and the IRAC data also
imply that the IRAC emission is primarily from H$_2$ lines in the IRAC
bands.  The overall appearance of the nebula in the 2.12 $\mu$m H$_2$ line
in the image presented hear and by \citet{speck02} also demonstrate the
correspondence between the near-IR and IRAC images of the nebula.  
Near-IR imaging of a cometary knot by \citet{huggins02} shows the H$_2$ 
emission concentrated at the face of the globule toward the central 
star, similar to the appearance in the 4.5 $\mu$m image.  The 
IRS spectra presented here confirm that there are no other significant
contributors to the IRAC band flux in the locations sampled.

There are, however, variations in the relative intensity of the IRAC
images across the nebula.  This is due to either changes in the relative
line strengths of the H$_2$ emission in the IRAC bands, or other minor
components of the emission. One example is the [\ion{Ar}{3}] line at 8.99
$\mu$m, which is detected in the ISO and IRS spectra, and could contribute
to the emission seen in the 8 $\mu$m image.  In the IRS spectra here,
the H$_2$ lines in the 8 $\mu$m band contribute approximately 25$\times$
the flux from the [\ion{Ar}{3}] line. For the IRAC 4.5 $\mu$m image, one
possible contributor is the Br$\alpha$ emission line at 4.05 $\mu$m. We
could find no published narrowband images of the Helix in this line.  
\citet{speck02} imaged the nebula in the 2.16 $\mu$m Br $\gamma$ line of H
I, and do not detect any emission to the upper limit of 7$\times 10^{-8}$
ergs s$^{-1}$ cm$^{-2}$ sr$^{-1}$.  The Br$\alpha$/Br$\gamma$ line ratio
will vary with extinction, observed ratios can range from $\sim$2.8 to 12
\citep[e.g.,][]{bunn95}.  For a typical cometary knot, the peak surface
brightness observed at the inner edge of the knot is approximately 0.5 MJy
sr$^{-1}$, with a 1 $\sigma$ noise of 0.007 MJy sr$^{-1}$. If the
Br$\alpha$/Br$\gamma$ ratio was greater than 12 in the Helix, it would be
approximately 1\% of the 4.5 $\mu$m band flux. Therefore the contribution
from the Br$\alpha$ line is unlikely to be significant in the knots.

\begin{figure}
\includegraphics[scale=0.5]{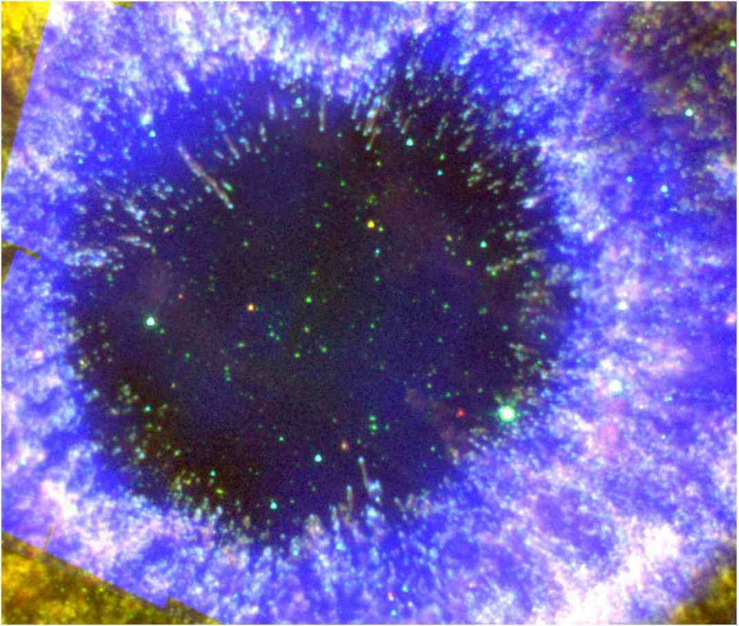}
\caption{
A three-color image of the central region of the Helix.  The IRAC 8 $\mu$m
image is red, 4.5 $\mu$m is green, and the ACS F658N image (H$\alpha$+[N 
II])
is blue.  The ACS image was smoothed to match the resolution of the IRAC images.
The ACS image does not completely cover the field, which is evident in the
corners of this image.
\label{n7293_f658_2_4_alignbig}
}

\end{figure}

Figure \ref{n7293_f658_2_4_alignbig} illustrates the relationship between
the emission detected by IRAC and the H$\alpha$ emission.  The IRAC 8
$\mu$m, IRAC 4.5 $\mu$m, and ACS F658N (H$\alpha$ + [\ion{N}{2}]) images
are plotted as red, green, and blue in this color image.  The ACS image
has been smoothed to match the IRAC 4.5 $\mu$m resolution.  The image
highlights the major differences between the H$\alpha$ and IRAC emission
-- in the inner region, the cometary knots are brightest in H$\alpha$ at
the tips.  Some tips are decidedly bluer, others appear a little green,
perhaps due to extinction that affects the H$\alpha$ to a greater degree.  
In the main ring, the emission differs drastically -- the H$\alpha$
emission is mostly smooth and seems to fill the whole region fairly
uniformly, whereas the IRAC emission is very clumpy.

\subsection{2.12 $\mu$m H$_2$ Image}

\begin{figure}
\includegraphics[scale=0.5]{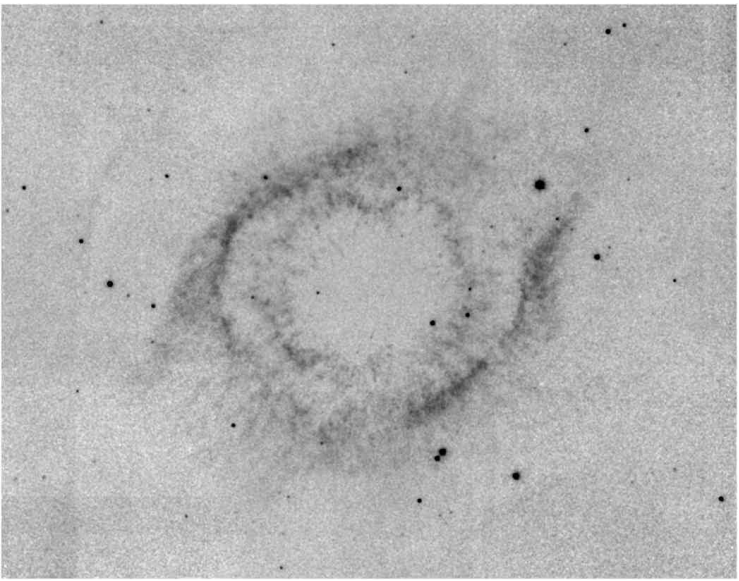}
\caption{
Narrowband H$_2$ image (2.12 $\mu$m) image of the Helix obtained
with QUIST.  North is up, and 
the image is approximately 29$\times$23 arcmin.  
\label{n7293_quist}
}
\end{figure}

The 2.12 $\mu$m H$_2$ image obtained from Mauna Kea is shown in Figure
\ref{n7293_quist}.  The image has lower spatial resolution and is not as
deep as the IRAC images, but the overall structure of the 2.12 $\mu$m
H$_2$ line is very similar to the emission in the IRAC images.  The image
is similar to the H$_2$ image of \citet{speck02}.  Here we have imaged a
larger area, and detect the faint radial rays extending from the ring to
the outer shell. The Northeast Arc is also detected.  Due to the lower
resolution, it is more difficult to see the individual cometary knots,
although the bright and better-separated cometary knot 428-900
\citep{odell97} and associated knots are visible.

\subsection{IRS Spectra}

\begin{figure}
\includegraphics[scale=0.45, angle=0]{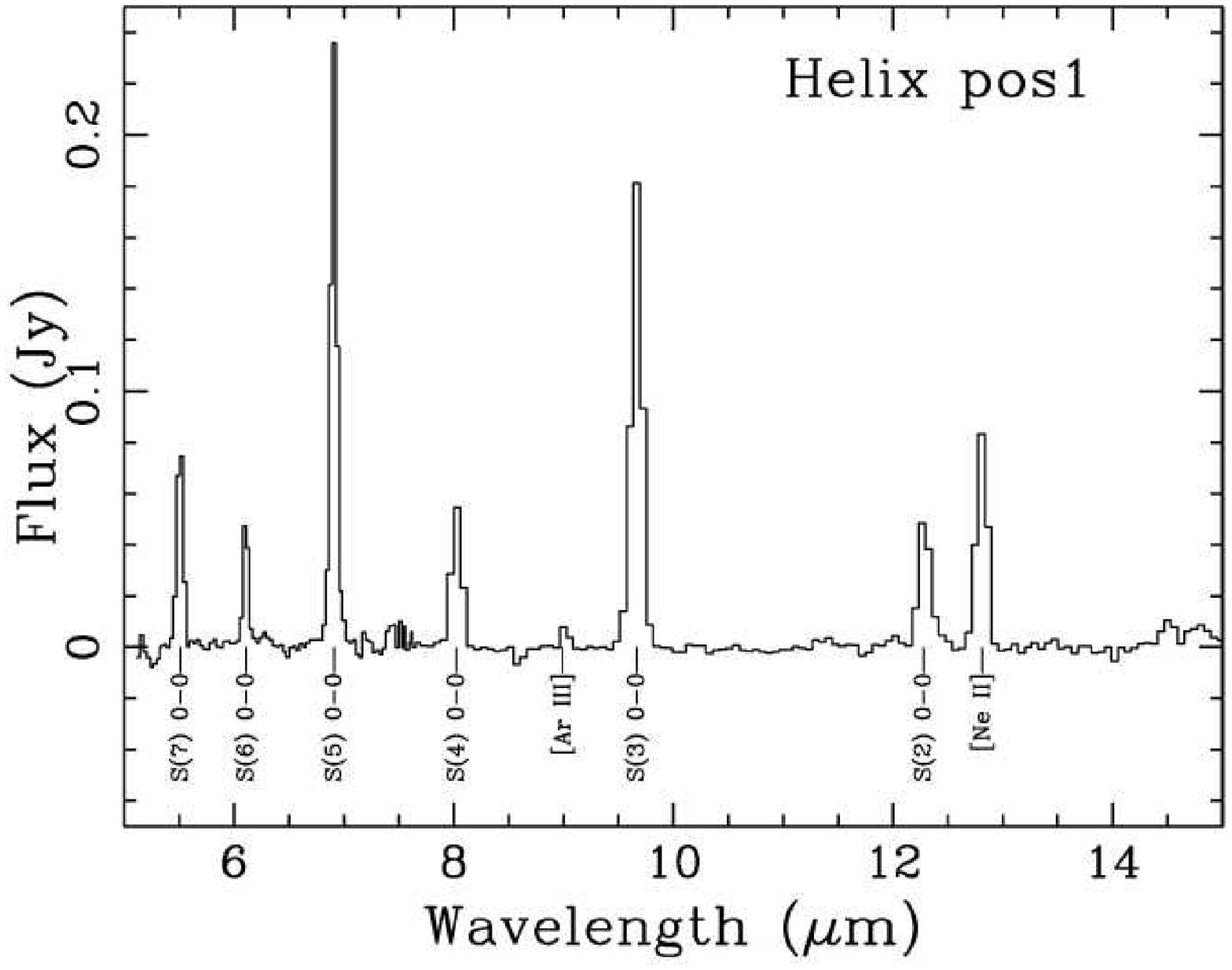}
\includegraphics[scale=0.45, angle=0]{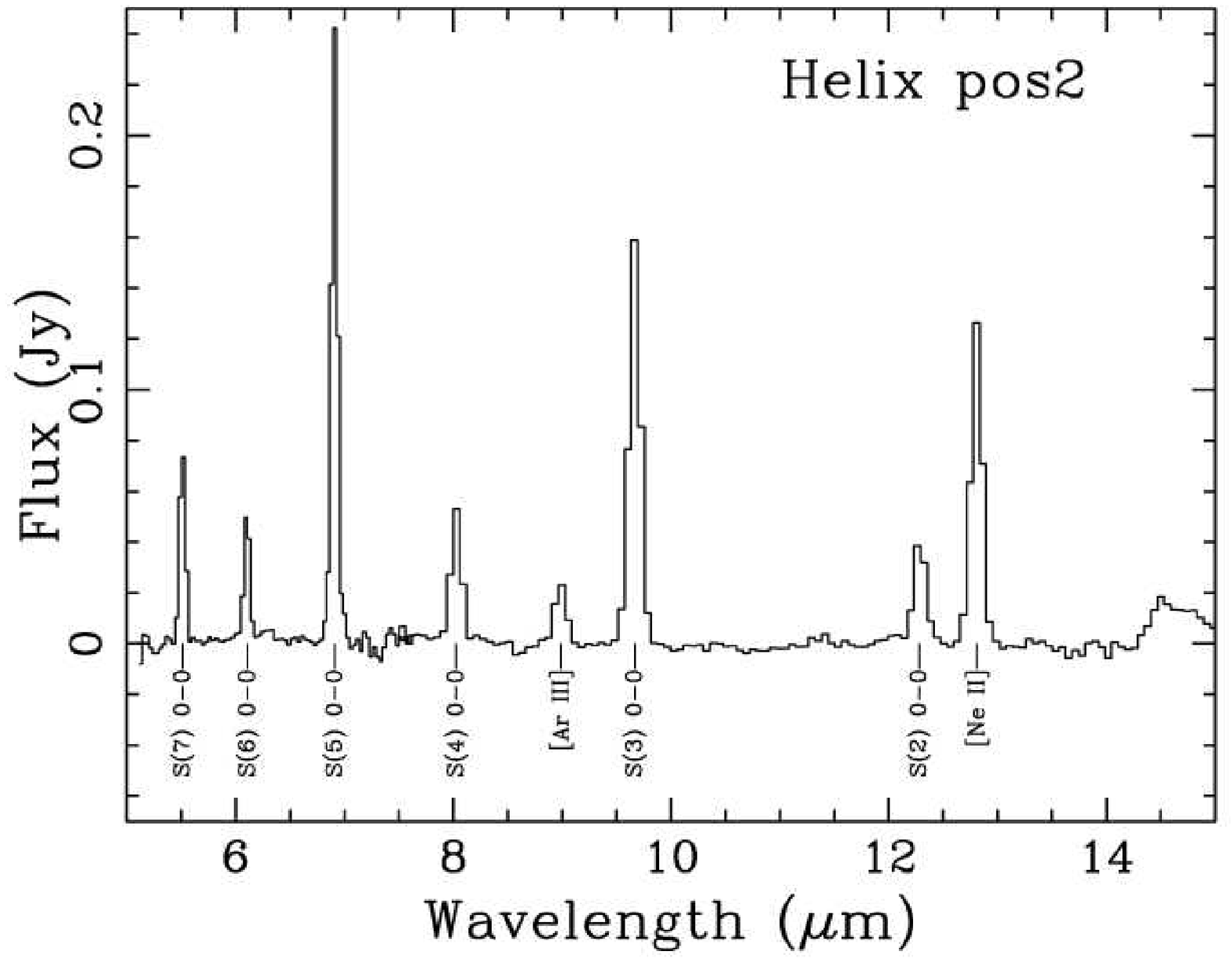}
\caption{
IRS spectra of the two positions in the Helix (see Figure \ref{IRAC_zoom}).  The spectra
from the "IRS OFF" position was subtracted before extracting these spectra.
\label{n7293pos1}}
\end{figure}

\begin{figure}
\includegraphics[scale=0.5]{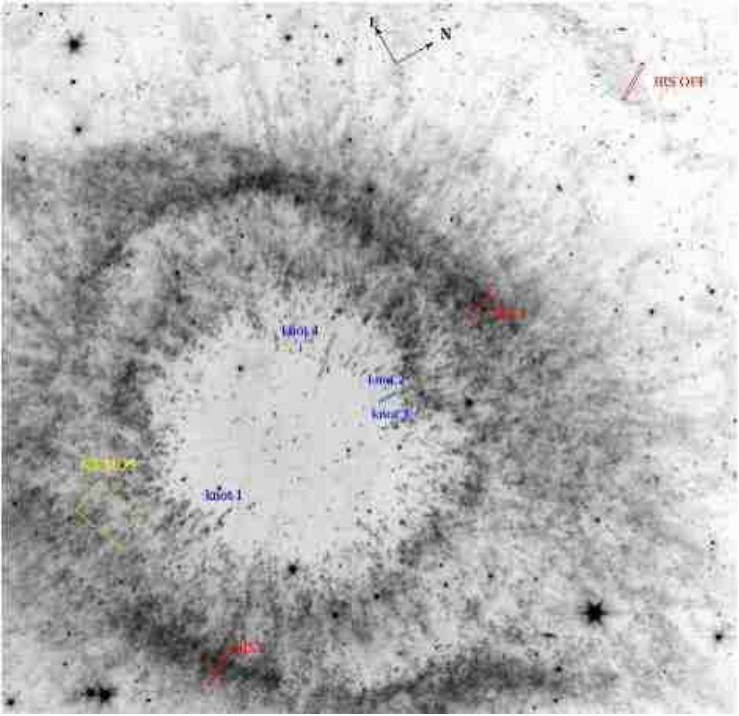}
\caption{
Location of the IRS slits and the knots that are discussed in the text, plotted
on the IRAC 4.5 $\mu$m image.\label{n7293_poslabel}
}
\end{figure}

Figure \ref{n7293pos1} shows the IRS spectra from the two locations
indicated in Figure \ref{n7293_poslabel}. The two locations are at
approximately the same radial distance and both in the main rings, so as
one might expect the observed spectra are very similar.  The IRS data used
for the background subtraction can be seen in Figure \ref{n7293_poslabel}
to fall across the faint outer arc, which may slightly affect the spectra
in Figure \ref{n7293pos1}.  However, assuming the emission in the arc is
due to H$_2$, it is a factor of 10 lower than the emission in the ring, so
it will have little effect on the line fluxes measured for the two ring
positions.

The dominant emission features from the nebula in this wavelength range
are from the pure rotational lines of molecular hydrogen.  The measured
line fluxes are given in Table~\ref{irslines}. These results are similar
to what was found by \citet{cox98} who found that the emission from 5 -
16.6 $\mu$m was dominated by H$_2$ line emission, with a small
contribution from the [\ion{Ar}{3}] 8.99 $\mu$m and [\ion{Ne}{2}] 12.81
$\mu$m lines.  They found the [\ion{Ne}{3}] 15.55 $\mu$m line to be
strong, but the forbidden Ne lines are outside of the IRAC wavelength
range.  Therefore, the emission in the IRAC 5.8 and 8 $\mu$m bands is
primary from the S(3) to S(7) lines of H$_2$.  The IRS spectra do not
sample the 3.1 -- 5.2 $\mu$m range of IRAC bands 1 and 2, so we do not
have a direct measurement of the emission components.  However, there are
H$_2$ lines present throughout the 3 - 5 $\mu$m range \citep[e.g.,
][]{black87} which are expected to be a major component of the emission in
bands 1 and 2 (see Section 4.2).

We compared the emission line flux in the IRS spectra to the IRAC flux
densities in bands 3 and 4.  Using the IRAC transmission curves
\citep{fazio04}, the expected flux densities at position 1 and 2 were
determined based on the measured IRS spectra, and are given in Table
\ref{irsflux}.  The comparable IRAC flux was estimated by summing the
emission in a 3\farcs6$\times$57\arcsec\ box centered at the IRS slit
position.  The IRS fluxes are slightly lower than the observed IRAC
fluxes, but given the uncertainties in the extended source calibration and
the way the fluxes were estimated, the measurements are consistent.  The
ratios between the two bands at each position are similar between the IRS
and IRAC measurements, indicating that the line ratios in the IRS spectra
are reasonably accurate.

\begin{table}
\begin{center}
\caption{Line Fluxes from the IRS Spectra\label{irslines}}
\begin{tabular}{crrr}
\\
\tableline\tableline
Line ID & Wavelength & Position 1 & Position 2 \\
 & ($\mu$m) & 10$^{-20}$ W cm$^{-2}$ &  10$^{-20}$ W cm$^{-2}$ \\
\tableline
H$_2$ (0,0) S(7) & 5.51116  &  5.85&  5.15\\
H$_2$ (0,0) S(6) & 6.10856  &  2.44& 2.73\\
H$_2$ (0,0) S(5) & 6.90952  &  10.4&  10.5\\
H$_2$ (0,0) S(4) & 8.02505  &  3.08& 2.91\\
$[$ \ion{Ar}{3} $]$ &  8.99138  & 0.29&1.37 \\
H$_2$ (0,0) S(3) &  9.66491   &  7.45 & 6.77 \\
H$_2$ (0,0) S(2) & 12.2786  &  1.37& 1.15\\
$[$ \ion{Ne}{2} $]$ & 12.8135 &  2.01 & 3.17 \\
H$_2$ (0,0) S(1) & 17.0348  &  2.38&\nodata \\
\tableline
\end{tabular}
\end{center}
\end{table}

\begin{table}
\begin{center}
\caption{IRAC Fluxes in the IRS Spectra Positions\label{irsflux}}
\begin{tabular}{crr}
\\
\tableline\tableline
 & Position 1 & Position 2 \\
\tableline
IRS 5.8 $\mu$m Flux (mJy) & 6.2 & 6.0  \\
IRS 8 $\mu$m Flux (mJy) & 8.3 & 8.2  \\
Ratio $F_{5.8 \mu m}/F_{8 \mu m}$ & 0.76 & 0.59\\
IRAC 3.6 $\mu$m Flux (mJy)& 1.5 & 1.7\\
IRAC 4.5 $\mu$m Flux (mJy)& 2.8 & 3.2\\
IRAC 5.8 $\mu$m Flux (mJy)& 5.9 & 6.0\\
IRAC 8 $\mu$m Flux (mJy)& 7.8 & 7.6 \\
Ratio $F_{5.8 \mu m}/F_{8 \mu m}$ & 0.76 & 0.78\\
\tableline
\end{tabular}
\end{center}
\end{table}

Absent from the spectrum is any trace of PAH emission.  \citep{cohen05}
showed that the 7.7 $\mu$m PAH feature is seen in PNe with nebular C/O
ratios of greater than 0.6. \citet{henry99} determined that the Helix had
an average nebular C/O ratio of 0.87, therefore one would expect PAH
emission.  \citet{cox98} argue that the nebula is carbon-rich, since 1)
molecular species such as CN, HCN, HNC, and HC0+ are comparable with the
abundances measured in carbon-rich nebulae, and 2) a high abundance of
neutral carbon measured near the western rim is only expected in
carbon-rich environments.  The lack of detected PAH emission could be a
sensitivity effect, since the nebula is very extended and the slit takes
in a small fraction of the nebula, compared to other measurements where
most or all of the nebula is sampled by the spectrograph beam.  
\citet{cohen05} note that this may be the case for two PN with large
spatial extent where they did not detect PAHs, however they also note that
the nebulae may be optically thin to ionizing radiation and no PDRs exist
in the PN.

\section{H$_2$ Line Emission}

\subsection{Modeling the H$_2$ emission observed at the IRS positions}

We modeled the H$_2$ line emission using the calculations of Michael Kaufman 
and Mark Wolfire for shocks and PDRs \citep{kaufman05,kaufman96,wolfire90}. The 
absence of any detectable lines from the higher vibrational states in our IRS 
spectra, in particular the normally strong 1-1 S(7) line, is convincing 
evidence that the emission is thoroughly dominated by shocks, although from 
our limit on the 1-1 S(7) line we cannot exclude a modest contribution from 
PDRs at a level below about 10\% of the shock contribution. We therefore have 
modeled the emission relying completely on C-shock excitation, and have done 
so successfully. We are able to fit the observed IRS line strengths to better 
than a few percent with a combination of three shocks with velocities of 20 
km s$^{-1}$, 35 km s$^{-1}$, and about 5 km s$^{-1}$. The dominant shock has 
a velocity of 20 km s$^{-1}$, with 75\% of the strongest feature (the 0-0 S(5) 
line) being produced in this shock. A 20 km s$^{-1}$ shock is unable to explain 
the relative strengths of the more highly excited 0-0 S(7) or 0-0 S(6) lines, 
and an additional component with $v$=35 km s$^{-1}$ was required to boost these 
lines. This faster shock provides about 25\% of the observed flux of the 
fiducial 0-0 S(5) line. 

The 0-0 S(1) line at 17 $\mu$m is considerably stronger than predicted by the 
combination of these two shocks, but because it is so easily excited 
(its upper state is at 171K) any modest shock with v $\sim$ 5 km s$^{-1}$ 
can excite this line without producing significant amounts of other line 
emission; in our best model 66\% of the 0-0 S(1) line comes from such a very 
slow shock, which contributes less than a few percent to the other lines. 
We found that for the higher velocity shock cases a preshock density of 
n=1$\times 10^{5}$ cm$^{-3}$ consistently gave a marginally better fit than 
did any other density, and we adopted it throughout; both lower densities 
tended to produce an over-intensity of 0-0 S(3) emission. For the slow shock 
component, the density was not a sensitive indicator. There is one additional 
line we can use in our modeling, the 1-0 S(1) line at 2.12 $\mu$m. Our best, 
3-velocity component shock model contributes only 1$\times 10^{-20}$ W cm$^{-2}$
to this line. From the ground-based image in the 1-0 S(1) line at 2.12 $\mu$m 
in Figure \ref{n7293_quist}, we estimated the flux in this line in the 
3\farcs6$\times$57\arcsec\  region of the IRS slit is 
2$\times 10^{-20}$ W cm$^{-2}$, considerably more than we can account for 
with this scenario. However a PDR contribution could be fully consistent with 
this measured flux. The limit to a PDR contribution from the limit in our 
IRS data, to the 1-1 S(7) line, is less than about 10\% of the shock 
contribution. A PDR with density of n=1$\times 10^{3}$ (insensitive to the 
local UV field) that contributes less than about 5\% to the pure rotational 
lines gives a 1-0 S(1) flux of 1.2$\times 10^{-20}$ W cm$^{-2}$. When this 
is added to the shock contribution, it can explain the observed emission 
as measured in the 2.12 $\mu$m image. We emphasize that while our model uses 
a three-component shock and rejects parameters that are significantly different, 
it is of course most probable that a wide range of shocks are present and 
our model simply fits them all to these parameters.  \citet{sternberg95}
have examined  cases with very dense ($10^6 cm^{-3}$) high UV-excited  
($G = 2\times10^5$) gas, but this density is unrealistically more than even 
the highest density estimate of \citet{meix05} of $10^4 - 10^5$ cm$^{-3}$ 
for the small knots in the Helix, and is inappropriate in the Helix case.

The measured fluxes in the lines enables us to estimate the filling factor 
for the shock in the IRS 3\farcs6$\times$57\arcsec\ beam. The dominant 20 
km s$^{-1}$  component has a filling factor of about 1.3$\times 10^{-3}$, 
while the 35 km s$^{-1}$ shock has a smaller factor of 7.4$\times 10^{-5}$.
The total luminosity in all the hydrogen lines, from the combined best model, 
is  3 $\times 10^{-3} L_{\sun}$ from this region of the Helix, 60\% of which is 
from the 20 km s$^{-1}$ component. Our IRAC Band 3 flux in this IRS region is 
about 3 mJy; the total IRAC Band 3 flux for the entire Helix is about 1000 
times greater, about 3 Jy. If the shock is everywhere similar to the region we 
have mapped, then the total H$_2$ line flux from the Helix can be scaled 
accordingly, for a total value of about 3$L_{\sun}$, about 5\% of the total 
luminosity of the Helix. This number is consistent with estimates from 
other authors as well \citep[e.g.,][]{cox98}. Together with our estimates for the filling factor,
and scaling with luminosity for the entire nebula as above, these numbers
imply a total mass in excited H$_2$ of a few Earth-masses.

From the best fit model we are able to calculate the corresponding fluxes 
in the IRAC bands by converting each of the observed line strengths into 
a flux density in the appropriate IRAC band using the IRAC instrumental 
response procedures as described in the IRAC Data Handbook (Section 5.2), 
as also described in \citet[][they note that the Handbook
Version 2.0 contains some significant errors in the Photometry and
Calibration section]{smith06}.
We have used the corrected handbook values 
(W. Reach, private communication).

\begin{figure}
\includegraphics[scale=0.5, angle=0]{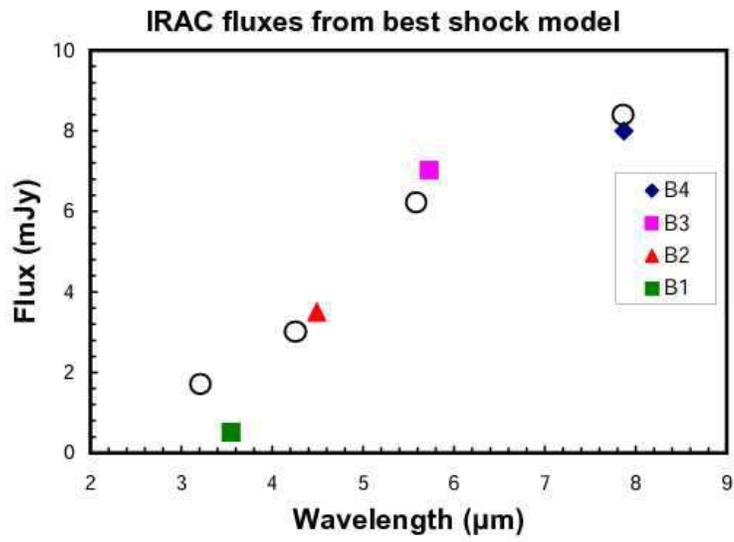}
\caption{
The colored symbols are the predicted contribution to the IRAC band fluxes
from H$_2$ line emission in the IRS slit region.  The circles are the
measured IRAC flux densities summed over the IRS slit.
\label{h2model}}
\end{figure}

Figure \ref{h2model} shows the net predicted contribution to the IRAC band 
fluxes from H$_2$ line emission in the IRS slit region; the circles are the 
measured IRAC flux densities summed over the IRS slit. Table \ref{band1_2flux}
shows the model fluxes from H$_2$ lines in IRAC bands 1 and 2. The results 
confirm a result noted by \citet{smith06} regarding the shock in DR21; 
namely, that in shocks the H$_2$ lines contribute significantly to the 
IRAC band fluxes. \citet{smith06} also report that in the shocked outflow of 
DR21, PAH emission (either at 6.2 $\mu$m or 7.7 $\mu$m) is absent. The 
Helix is even more dramatic than DR21 in that our spectra shows a very 
low continuum level consistent with zero at a level of about 
$3\times10^{-20}$ W cm$^{-2}$.  \citet{cohen05} found PAH emission in 
17 out of 43 PNe they studied with ISO, and concluded that as a 
general rule when the C/O ratio was less than about 0.6 there was no PAH 
emission; they do report, however, the exception of NGC 6720 which has 
a C/O ratio of 0.62 yet no detected PAH emission.  Our non-detection of 
PAH emission in the Helix is consistent with the broad trends Cohen and 
Barlow report.  As a result, the H$_2$ lines (as seen in Figure \ref{h2model}) 
contribute more than 90\% of the flux in the IRAC bands 2, 3 and 4. 

\begin{table}
\begin{center}
\caption{Molecular Hydrogen Model Fluxes in IRAC Bands 1 and 2\label{band1_2flux}}
\begin{tabular}{lrr}
\\
\tableline\tableline
Line ID & Wavelength & Flux Density \\ 
 & ($\mu$m) & (Jy) 
\\
\tableline
Band 1 (3.6 $\mu$m):\\
2-1 O(9) & 4.884 \\
*0-0 S(9) & 4.694 & .52mJy \\
1-0 O(9) & 4.576 \\
2-1 O(8) & 4.438 \\
*0-0 S(10) & 4.410 & .012mJy \\
0-0 S(11) & 4.181 \\
1-0 O(8) & 4.163 \\
2-1 O(7) & 4.054 \\
0-0 S(12) & 3.996 \\
 & total: & .54mJy \\
 &  \\
 &  \\
Band 2 (4.5 $\mu$m):\\
0-0 S(13) & 3.846 \\
*1-0 O(7) & 3.807 & .058mJy \\
2-1 O(6) & 3.724 \\
0-0 S(14) & 3.724 \\
0-0 S(15) & 3.626 \\
0-0 S(16) & 3.547 \\
*1-0 O(6) & 3.501 & .053mJy \\
0-0 S(17) & 3.485 \\
0-0 S(18) & 3.438 \\
2-1 O(5) & 3.438 \\
*1-0 O(5) & 3.235 & .31mJy \\
2-1 O(4) & 3.190 \\
 \\
 & total: & .42mJy \\
\tableline
\end{tabular}
\end{center}
\end{table}

Our limited spectral data on the Helix do not allow us to determine the 
reason for the absence of PAH emission here. In the case of the DR21 outflow, 
the possibilities considered for the absence of PAH were shock or uv 
depletion of the material, or the absence of a suitable PDR environment. 
\citet{cox98} studied the Helix and its H$_2$ lines with the ISOCAM CVF; 
they also conclude that PDR excitation in the Helix is inadequate to explain 
the H$_2$ lines they measured. They consider C- and J-shocks, but argue that 
neither of these scenarios offers a convincing answer either, C-shocks 
because of the weak magnetic fields thought to be present in the in the Helix, 
and non-dissociative J-shocks because they require lower densities and/or 
temperature than are inferred. Finally, they note that there is no strong 
evidence in the Helix for a stellar wind capable of generating such shocks. 
The fact that the Kaufmann, Wolfire and Hollenbach shock models fit so 
precisely our observed line fluxes, at least in the combination we describe 
above, suggests to us that we must reexamine all of these assumptions about 
conditions in the Helix Nebula. In particular, our data indicate that there 
are (or were) 
strong winds (or other shock-producing mechanisms) present in this 
planetary, and furthermore that there is enough magnetic field in the outer 
shell to prevent the higher velocity shocks from dissociating the molecules.
High velocity gas is well known to exist in the Helix.  \citet{young99}
mapped the molecular envelope in CO 2-1, and report gas motions of more than 
50 km sec$^{-1}$, more than adequate to shock the H$_2$ in our models. 
In their study of the origin and nature of the Helix structure, they conclude 
that there is evidence for directed bipolar flow in the early stages of 
development of the Helix, and that or related activity could produce the 
shocks we apparently see in H$_2$. 

\subsection{Variation of the H$_2$ emission in the knots and nebula} 

As noted earlier, the relative strength of the emission in the IRAC bands
varies as a function of distance from the central star, and also on small
scales as in the cometary knot tips and the inner edges of the structures
in the ring. Our IRS spectra only cover a tiny portion of the nebulosity,
and these locations were chosen because of the bright knots known to be
there; therefore some caution should be exercised when extrapolating our
spectral conclusions to the entire region.  Nevertheless, HST and radio
studies do not find any evidence that the area of our spectral study is
dramatically different in kind from other spots in the Helix, and it is
useful to see if a consistent picture can emerge from the IRAC images
alone with this proviso. As shown in \ref{n7293_mags}, 
the tips of the knots have
the highest 4.5/8.0 $\mu$m emission ratio, and the value in the main ring and
outward decreases as the radius increases. The 3.6/4.5 $\mu$m ratio remains
relatively constant over the same radius range, increasing only slightly
at higher radius. The H$_2$ models indicate that this trend can be matched by
a decreasing shock velocity and density as the radius increases. The
small-scale color differences could be a result of a higher PDR
contribution to the H$_2$ emission on the inner edges of the clumps.

\section{Cometary knots and clumps}

\subsection{Enhanced Structure images}

\begin{figure}
\includegraphics[scale=0.9]{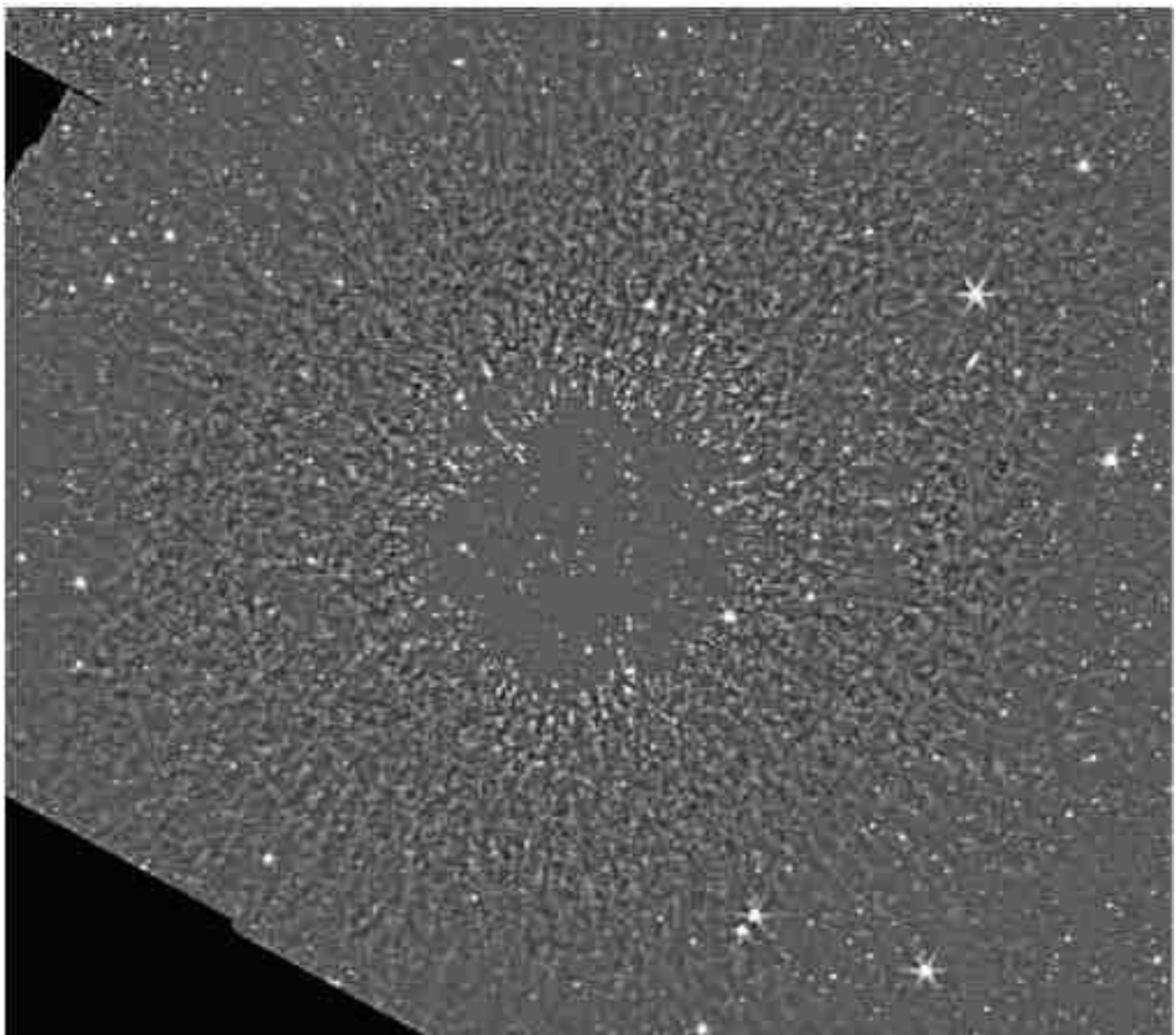}
\caption{
IRAC images with the small-scale structure enhanced by taking the ratio of
the original image to its local median, as described in Section 4.1. N is
at the top, and the field shown is approximately 45$\times$23 arcmin. a)
4.5 $\mu$m image.
\label{n7293_ch2med51x51ratio}}
\end{figure}

\begin{figure}
\includegraphics[scale=0.9]{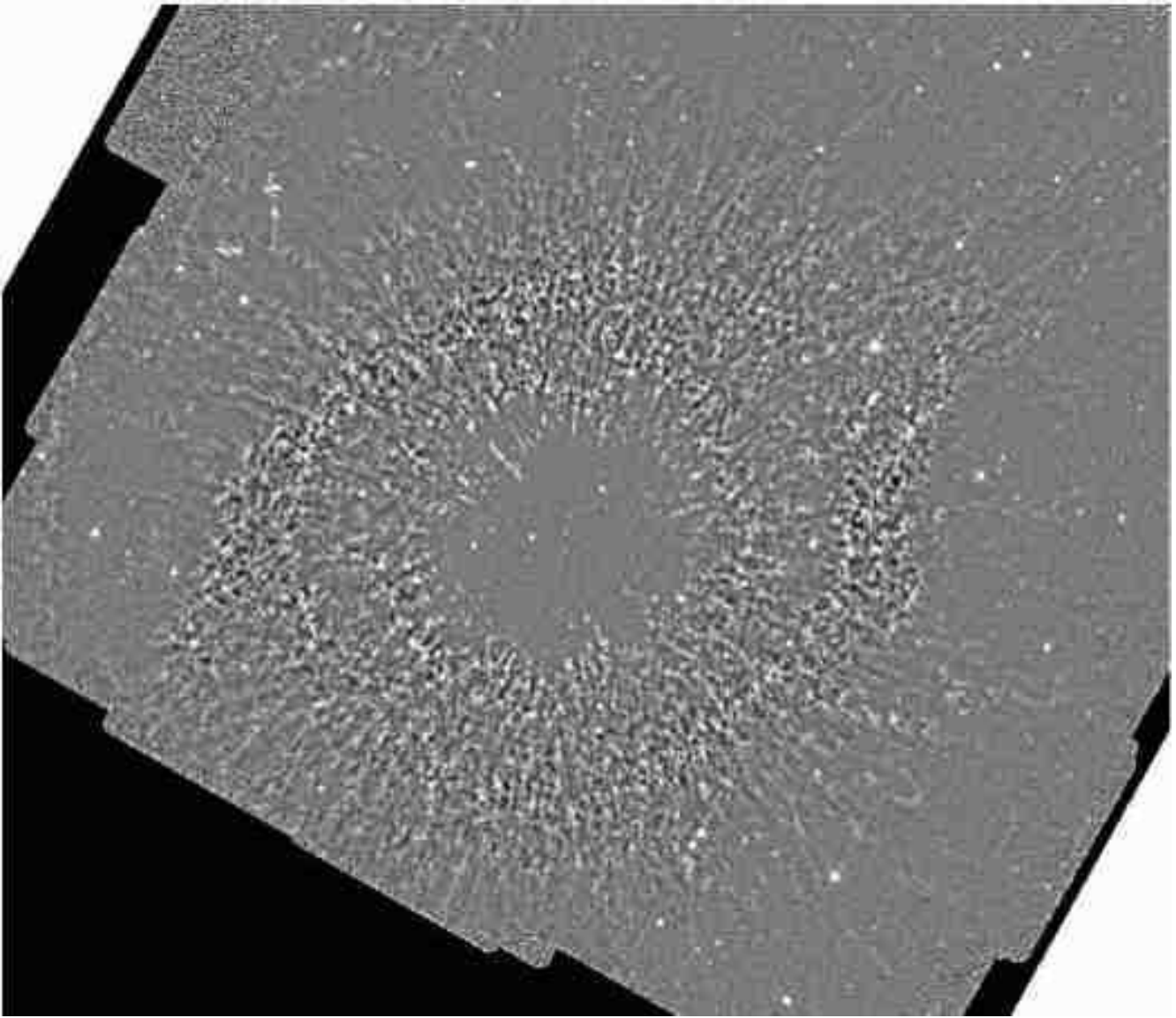}
\caption{same as Figure \ref{n7293_ch2med51x51ratio}, but the  
8 $\mu$m image. 
\label{n7293_ch4med51x51ratio}}
\end{figure}

\begin{figure}
\includegraphics[scale=0.9]{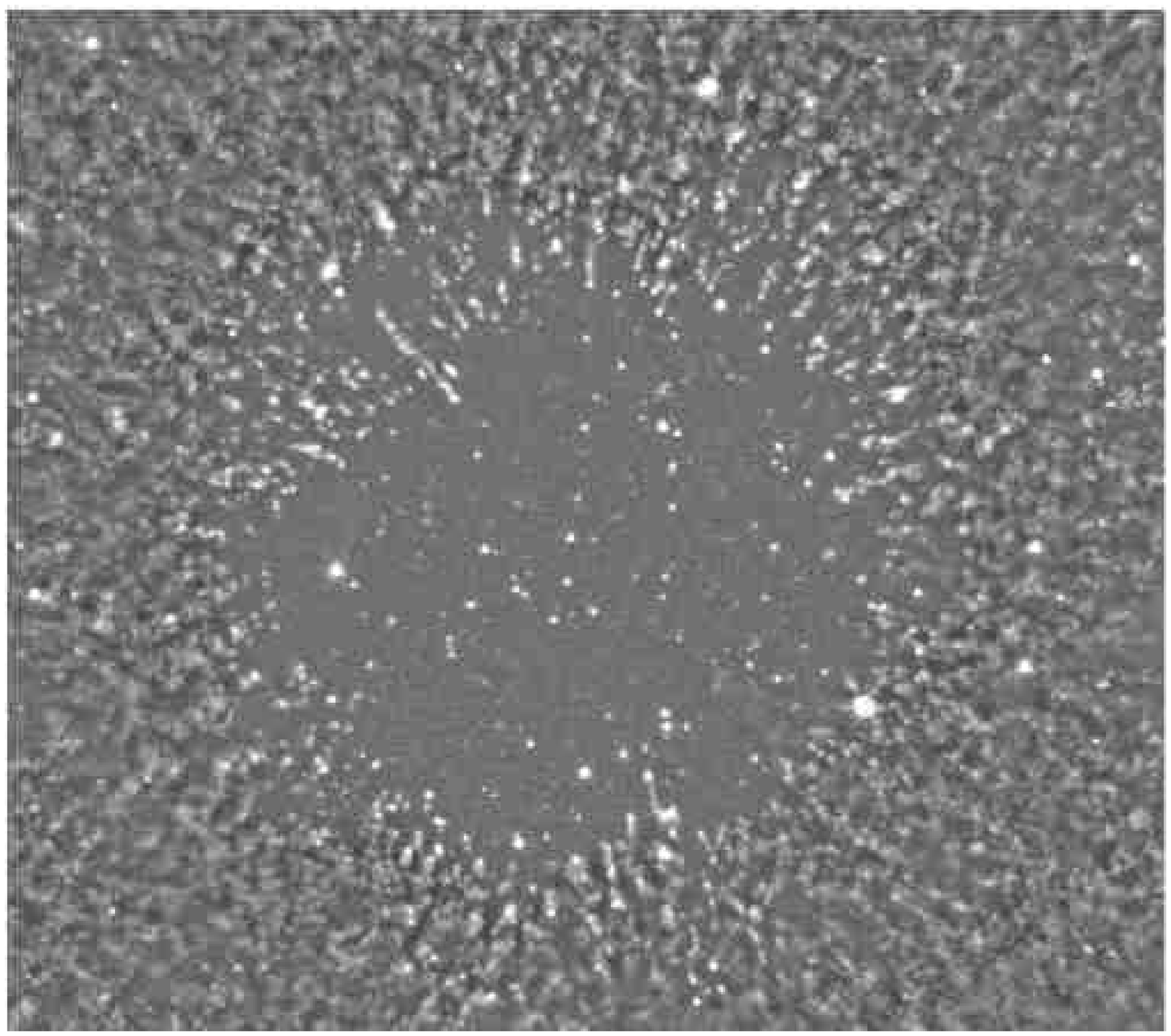}
\caption{
Same as Figure \ref{n7293_ch2med51x51ratio}, but showing the inner 10$\times$9 
arcmin region.
\label{n7293_ch2med51x51ratiozoom}
}
\end{figure}

\begin{figure}
\includegraphics[scale=0.9]{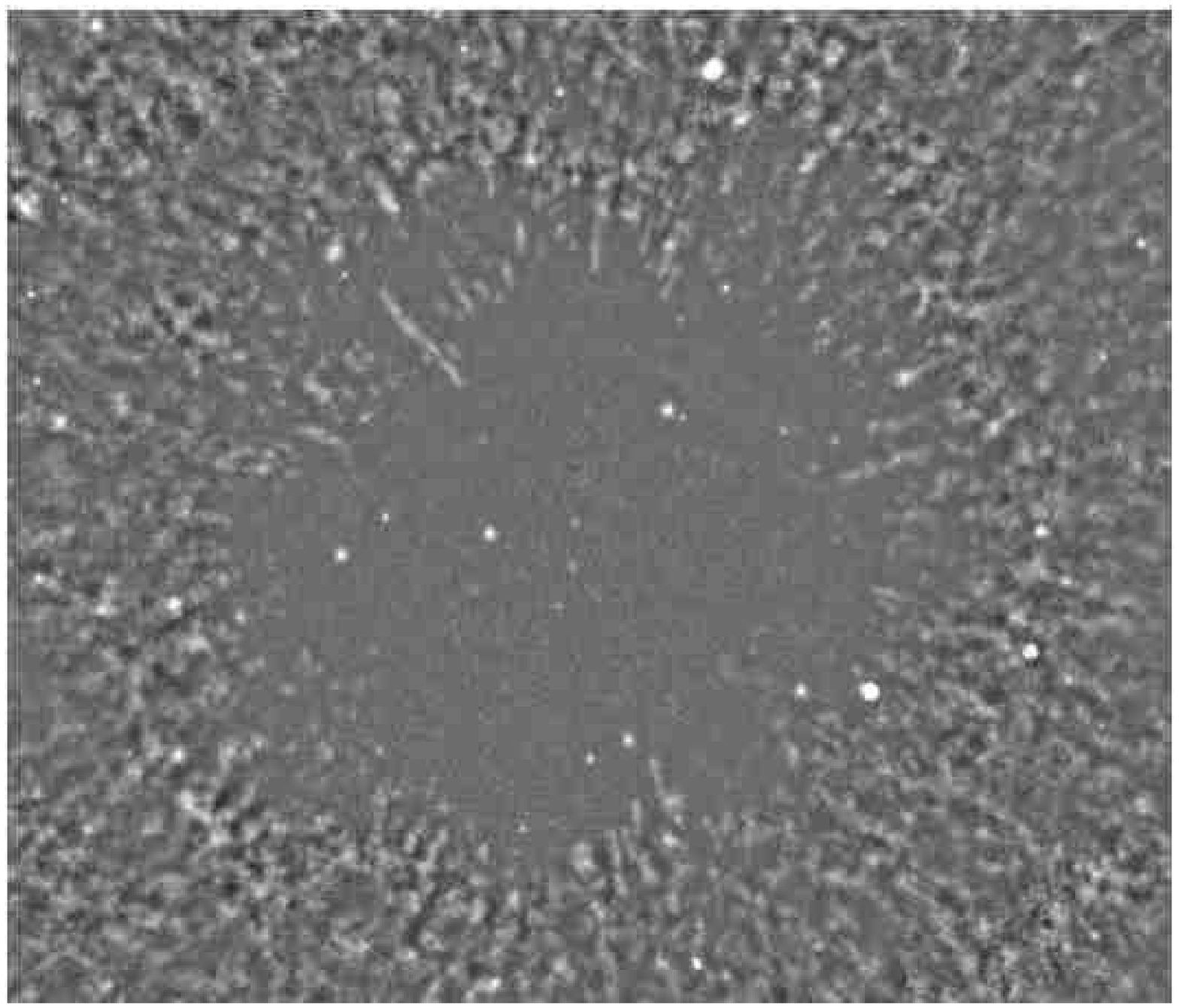}
\caption{
Same as Figure \ref{n7293_ch4med51x51ratio}, but showing the inner 10$\times$9 
arcmin region.
\label{n7293_ch4med51x51ratiozoom}
}
\end{figure}

\begin{figure}
\includegraphics[scale=0.85]{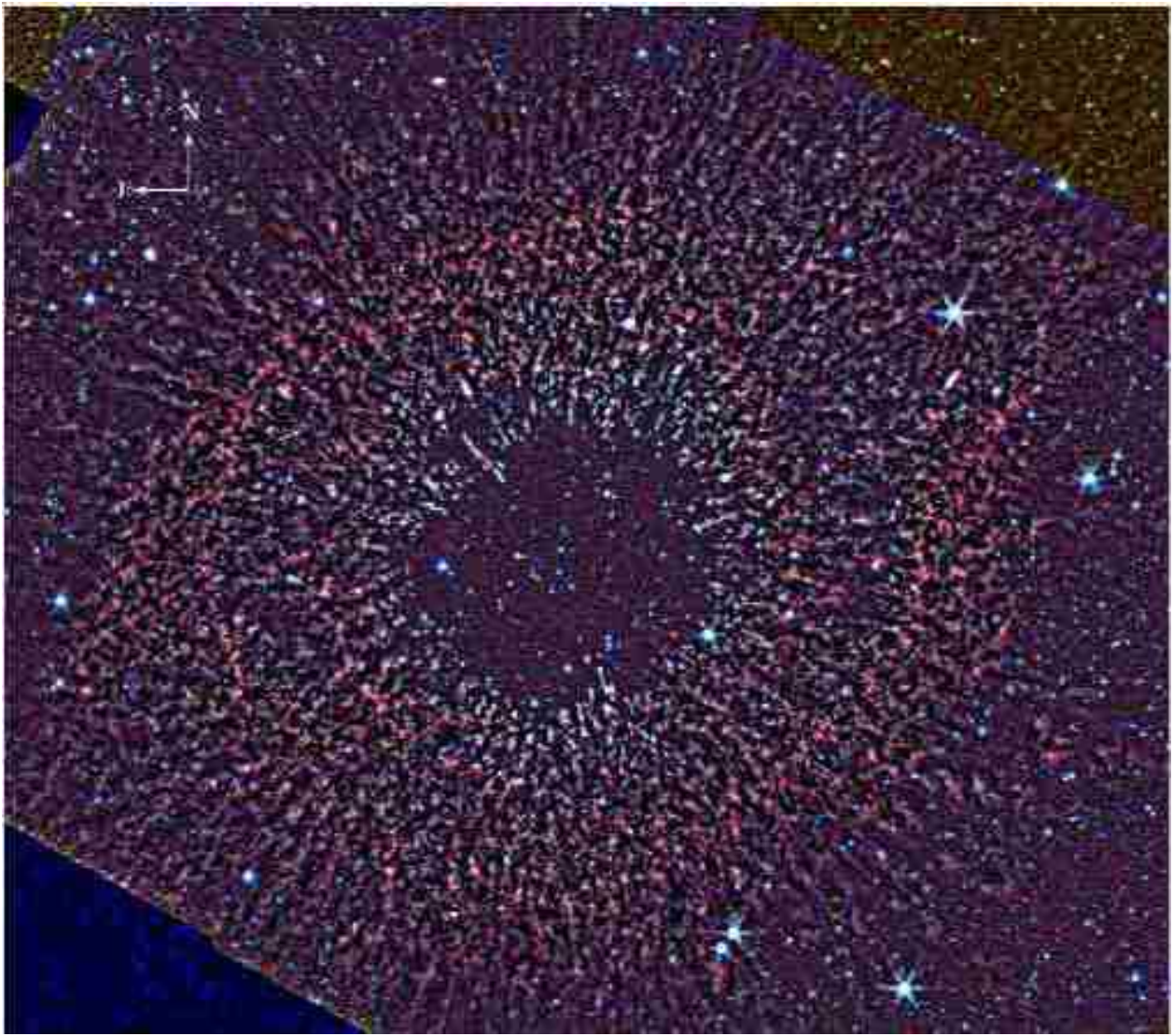}
\caption{
IRAC 3 color image (the 8, 4.5, and 3.6 $\mu$m images mapped to RGB)
using the structure-enhanced images as described in Section 4.1. 
\label{n7293_124_medratios}
}
\end{figure}

\begin{figure}
\includegraphics[scale=0.85]{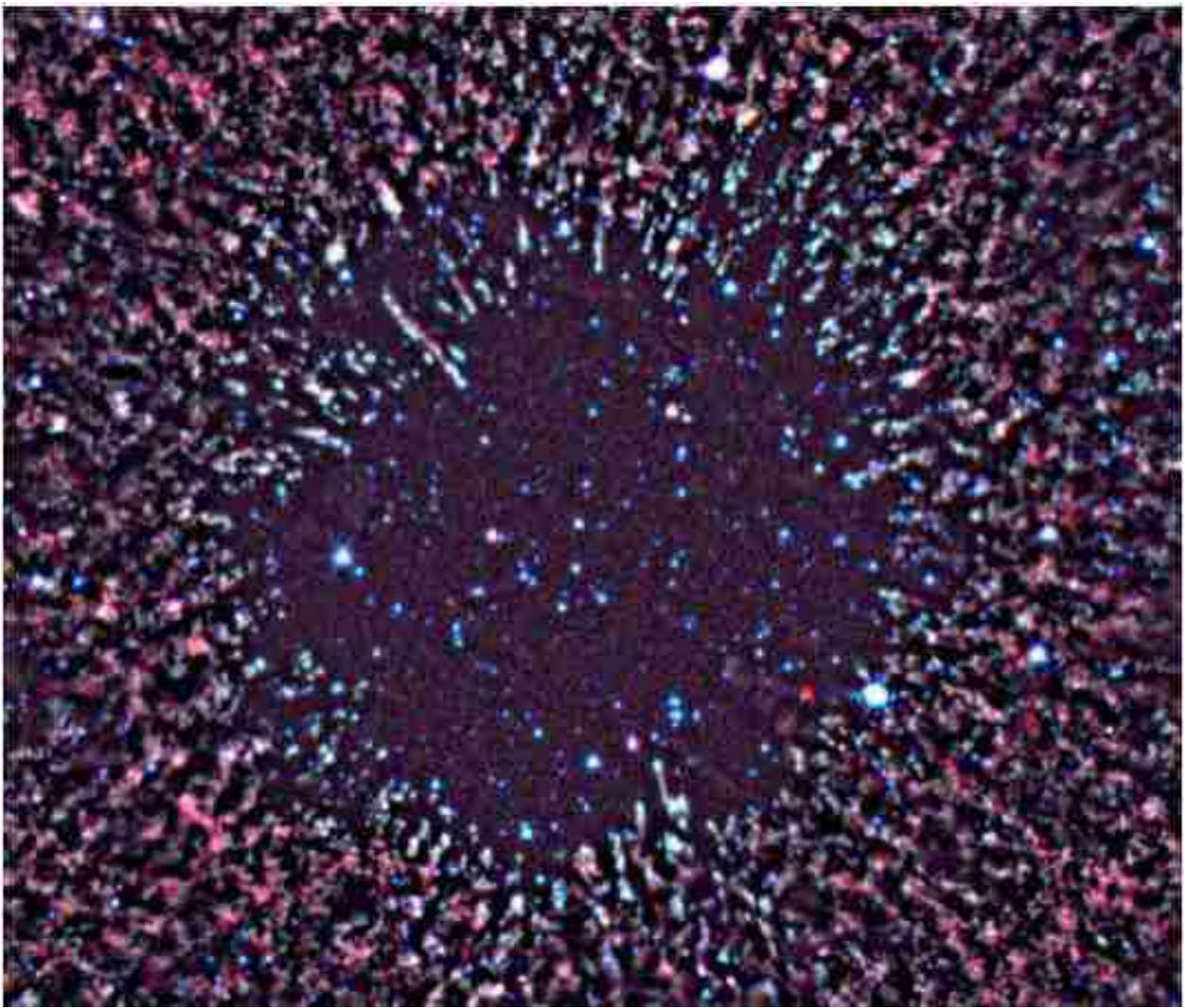}
\caption{
Three-color image (the 8 and 4.5 $\mu$m IRAC images and the ACS F658N image
are mapped to RGB) of the central region of the Helix.  This image used the 
structure-enhanced versions of the images as described in Section 4.1.
\label{n7293_h658_24_medratiosr}
}
\end{figure}

In order to show the small-scale structure more clearly, we processed the
IRAC images by dividing the image by a median-smoothed version of the same
image using a square kernel of 20\farcs4 in size, similar to what was
done by \citet{odell04} for their [\ion{O}{3}] image.  The results are
shown in Figures \ref{n7293_ch2med51x51ratio} --
\ref{n7293_h658_24_medratiosr}. Figures \ref{n7293_ch2med51x51ratio} and
\ref{n7293_ch4med51x51ratio} show the processed 4.5 and 8 $\mu$m images,
and Figures \ref{n7293_ch2med51x51ratiozoom} and
\ref{n7293_ch4med51x51ratiozoom} shows the inner
10\arcmin$\times$9\arcmin\ region.  Outside of the innermost region, the
appearance of the nebula is clumpy on small scales, lacking the long tails
of the cometary knots of the inner region.  The size and structure of the
clumps is fairly uniform across the nebula, and the long radial rays are
seen only in the outer regions, outside of the main ring.  The images show
that the H$_2$ structure observed in the NICMOS images of \citet{meix05}
in Figure \ref{ch2_nicmos} are representative of most of the main ring
region.

Color versions of the enhanced structure images are shown in Figures
\ref{n7293_124_medratios} and \ref{n7293_h658_24_medratiosr}. Figure
\ref{n7293_124_medratios} is a 3-color image of the 8, 4.5, and
3.6 $\mu$m IRAC enhanced structure images mapped to red, green, and blue,
respectively.  The color difference between the inner blue-green region of
the cometary knots and the outer redder clumps is visible here.

Figure \ref{n7293_124_medratios} shows the IRAC 8 and 4.5 $\mu$m images
mapped to red and green, and the ACS F658N image mapped to blue.  The
structure-enhanced versions of the images were used.  Two important
characteristics of the emission are shown in this image.  First, as seen
before, the color structure of the cometary knots are readily visible,
with their blue-green tips and red tails.  The other feature visible in
this image is that there are bright blue emission regions in many of the
locations where there is a minimum of IRAC emission.  This is a result of
the different sources of the emission in these bands -- the emission in
the ACS F658N filter is primarily from the H$\alpha$ and [\ion{N}{2}]
lines in the ionized regions of the nebula, whereas the IRAC structure is
primarily from the H$_2$ lines at the interface between the ionized
regions and the molecular material.  This shows that the two emission
components are fairly well-mixed on larger spatial scales, but on smaller
scales the emission is spatially segregated.  The H$_2$ has a clumpy
distribution, and the emission from the ionized gas is strongest in the
voids between the clumps.  This is consistent with either the gas not
being present near the H$_2$ clumps, or being shielded from the ionizing
radiation from the central star in the shadow of the clumps.

\subsection{Structure of the knots }

\begin{figure}
\includegraphics[scale=0.92]{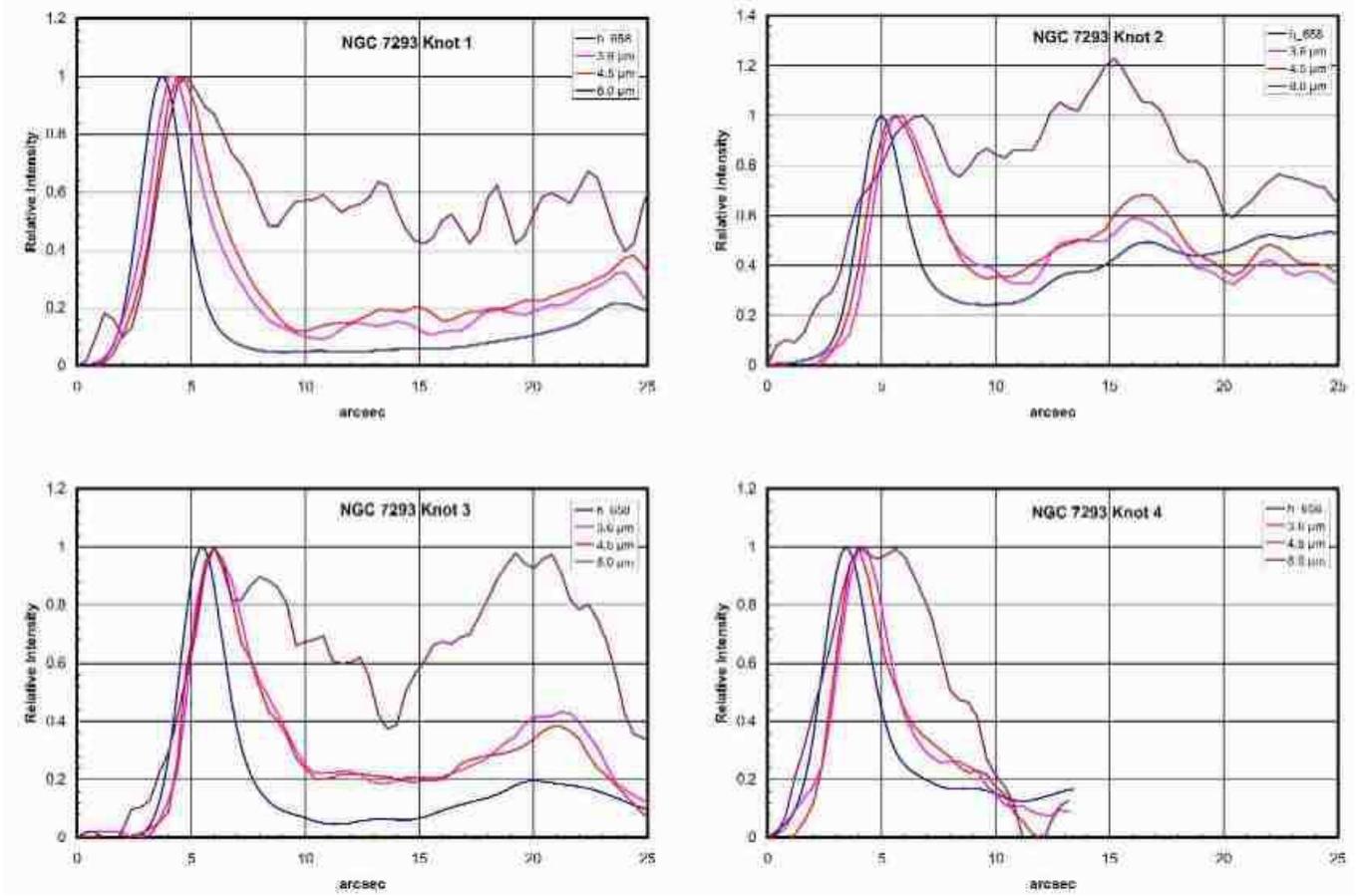}
\caption{Plots showing the profiles of the emission through each of four 
separate knots.  In each plot, the knot profiles from the 0.658, 3.6, 4.5, 
and 8 $\mu$m images are shown. 
The profile was extracted along a vector aligned in the radial 
direction, plotted 
in the inner to outer direction (see Figure \ref{IRAC_zoom}
and Table 2 for the location of
the knots). The width of the extracted region was 2\arcsec.
\label{knots}
}
\end{figure}

Figure \ref{knots} shows profiles of the emission through four cometary
knots.  The knots examined are indicated in Figure \ref{n7293_poslabel},
and the identifications and positions are given in Table \ref{knottab}.  
The profile width was 2\arcsec. The 0.658 $\mu$m (H$\alpha$+[\ion{N}{2}])
image was convolved with a Gaussian to match the spatial resolution of the
IRAC images.  The 0.658 $\mu$m emission peak appears at the inner edge of
the knots, and IRAC bands 1,2, and 4 peak about an arcsec or more behind
the optical line emission in each case.  The emission drops off rapidly in
all bands except for the 8 $\mu$m band, which drops off to a plateau
that extends the length of the tail.  The IRAC 3.6 and 4.5 $\mu$m bands
drop off slightly less than the optical emission, but roughly follow the
0.658 $\mu$m profile.

The structure of the H$_2$ emission in the cometary knots in the IRAC
images differs from that reported by \citet{walsh03} who report that in
their 2.12 $\mu$m imaging of the cometary knots with 1\farcs2 seeing,
the H$_2$ emission is seen only in the low ionization region facing the
central star, but not in the cores of the knots. They do note that H$_2$
was seen in some of the tails. With IRAC, the H$_2$ emission is seen all
along the knot, although with IRAC's resolution, the emission from the rim
of the knot would fill in the neutral core and we would not resolve an
emission-free region.  However, the spatial distribution of the 2.12
$\mu$m emission reported by \citet{walsh03} is consistent with the trend
seen in the IRAC data where at shorter wavelengths, more of the emission
is concentrated in the tip of the knot, and less along the length of the
cometary tail.

\subsection{Number of knots}

There have been several estimates of the number of knots in the Helix,
with \citet{meix05} recently updating the estimate based on their
NICMOS observations.  They counted the number of knots in the field of
view and assumed the same statistics over the area of the main ring,
concluding that there are a total of $\sim$23,000 knots.  Because we have
imaged the entire ring with IRAC, we can more directly estimate the number
of knots without extrapolating.  We used the IRAC 8 $\mu$m image for the
estimate because it is the least sensitive to optical depth through the
nebula, and provides the most accurate total flux from the knots.  Due to
the limited resolution and large number of overlapping knots in the main
ring, it is impossible to individually count the knots from the image.  
The number of knots can be estimated by determining the average flux from
one knot, and then dividing the total flux by the average to obtain the
number of knots.  We examined individual knots inside the main ring to
determine the average flux.  The HST/ACS image was used to identify the
knots and their boundaries, and the average fluxes for a sample of
30 knots were determined.  The area used for the average included
the bright heads of the knots as well as emission from the fainter tails.
The knot fluxes varied over a factor of two (from $\sim$ 240 -- 480
$\mu$Jy/knot), resulting in an estimate of 20,000 -- 40,000 knots in the
main ring.  Adding to the uncertainty is the fact that the appearance of
the knots changes as a function of radius.  If the change is not just an
evolution of the morphology but a change in the total H$_2$ emission as a
result of different mass or excitation, then this estimate of the number
of knots will be affected.

We cannot directly estimate the total mass of the knots, since the
H$_2$ emission detected by IRAC
is from the surface of the knots and comprises a relatively
small fraction of the mass.  \citet{meaburn92}, \citet{odell96}, and
\citet{huggins02} estimate the total mass of the knots in the range of
1 -- 2 $\times 10^{-5} M_\sun$.  Adopting a mass of 1.5$\times 10^{-5}$,
the total mass
of the knots would be in the range of 0.3 - 0.6 $M_\sun$. This is comparable
to the total ionized mass estimated by \citet{henry99} of 0.3 $M_\sun$,
and in agreement with the result of \citet{meix05}.

\section{Halo Structures}

Outside of the main rings, the character of the emission changes.  The
images in Figures \ref{4colfull}, \ref{n7293_ch2med51x51ratio}, and
\ref{n7293_h658_24_medratiosr} show that the region between the main ring
and the Northeast-arc (in the nomenclature of \citet{odell04}; hereafter
NE-arc) is filled with long radial rays and large arc-shaped structures,
in contrast to the main ring where the compact knots dominate.  The rays
in the halo also differ from the cometary knots in that they don't have
bright inward-facing tips, and the features in the halo are much broader.
The emission extends to the edges of the IRAC images, beyond the shell
that the NE-arc is the brightest feature.  In the 5.8 and 8 $\mu$m
images in particular, one can see the rays extending beyond the NE-arc
ring in the east and north. In the eastern corner of the 5.8 $\mu$m image,
a rim of another outer shell seems to be visible about where the southeast
plume terminates.  We have planned further IRAC observations to determine
the extent of this outer halo.

\subsection{Northeast Arc (NE-arc)}

\begin{figure}
\includegraphics[scale=0.92]{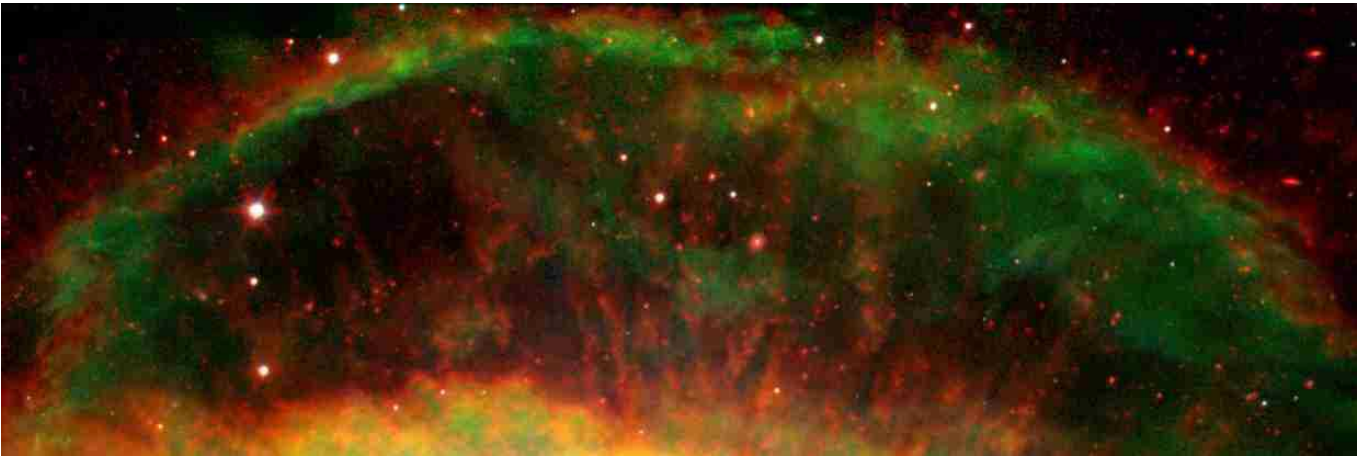}
\caption{
Combined Spitzer (4.5 and 8 $\mu$m) and HST 
H$\alpha$ image \& [\ion{O}{3}] image shown as red-green-blue,
respectively (full image available at
http://www.spitzer.caltech.edu/Media/releases/ssc2006-01/ssc2006-01b.shtml).
North is 60.9 deg CW from up.  The image shows clearly that the H$_2$ 
emission is located in a shell just outside of the H$\alpha$ emission. 
\label{Opt_IR_halo}
}
\end{figure}

Figure \ref{Opt_IR_halo} shows a comparison between the infrared and
optical emission of the NE-arc.  In this color image, green is 
H$\alpha$ + [\ion{N}{2}]
at 6369 \AA, and red is a combination of the IRAC 4.5 and 8 $\mu$m
bands.  The characteristics of the emission is similar in the optical and
infrared, except that the IR emission, primarily from H$_2$, is located in
a shell adjacent to and outside of the optical emission.  The structure of
the shell differs greatly from the main ring which is about half as far
from the central star.  Instead of being very clumpy, at or below the
resolution of the IR images, the features in the arc are more extended,
with a minimum width or size of 10\arcsec.  The emission is in a
relatively thin shell, compared to the main rings which are radially
thicker. The morphology of the arc and the relative locations of the
H$\alpha$ and molecular emission suggest that this arc is a PDR viewed
nearly edge-on.

The separation of the ionized gas and H$_2$ emission in the NE-arc
is markedly different than in the main ring.  In the ring, the clumps of
emission seem to coexist with the ionized gas.  What structure that exists
in the H$\alpha$ emission seems to be anti-correlated with the H$_2$
clumps, but not segregated radially.  It is difficult to envision a
scenario where complicated structures like those in the main ring and
cometary knots could evolve to a simpler shell structure like that in the
Arc.  Therefore it would seem that the mass ejection episode that created
the main rings was quite different than the one that created the outer
shell.  There are several examples of other PNe that also have more
spherically symmetric or simpler outer halos, but complicated inner
structure.  For example, NGC 6720 has a clumpy asymmetric inner main ring
surrounded by a nearly circular outer arc \citep{speck02}.  NGC 6543 has a
complicated inner structure surrounded by spherical shells
\citep{mitchell05}.

\subsection{Radial Rays}

The region between the main ring and the outermost arcs are filled with
small arcs and radial rays. These features are visible in all of the
bands, but are most pronounced in the 5.8 and 8 $\mu$m images in Figure
\ref{IRAC_all}.  Some of the small arcs could be parts of the same shell
as the NE-arc seen in projection on the sky at smaller radial
distances.  However, the rays seem to be different in character, with
their long dimension oriented radially from the central star, and
extending from the main rings to the outer arcs and beyond.

The rays are also visible in optical images.  The [\ion{O}{3}] images of
\citet{odell04} show rays as well (their Figs. 13 and 16), but primarily
in and just outside of the main ring.  The rays are not as visible in the
main ring in the IRAC images, in part because of the greater number of
emission clumps in the IR data.  Outside of the main ring, the rays are
visible in the IRAC data but the brightness in the [\ion{O}{3}] drops
rapidly.  Where the IR and optical rays overlap, there is little or no
correspondence to the positions of the optical and IR rays; in fact, they
seem to be if anything anti-correlated.  This implies that the IR rays are
in the shadow of clumps in the main ring, and the optical rays are where
the light from the central star shines through holes between the clumps.  
However, it is difficult to trace the rays to any one clump or group of
clumps because of their density;  if one traces a line inward from the IR
ray, one passes through many clumps.

Recent optical imaging by \citet{meaburn05} in the H$\alpha$ +
[\ion{N}{2}] lines show faint radial rays in the inner region, at smaller
radii than the cometary knots.  The IRAC images do not show any trace of
these structures; the inner region has only faint diffuse emission that is
on the order of 1\arcmin\ in size.  These could be wisps of halo emission
that are in front of or behind the main nebula and seen in projection
close to the central star.

\section{Conclusions}

We have presented IRAC images and IRS spectra of the Helix PN. The
emission from the nebula is dominated by the pure rotational lines of
H$_2$ with a smaller contribution from forbidden line emission such as
[\ion{Ar}{3}] in the ionized region.  The H$_2$ emission is consistent
with models of shock excitation, with an approximately 10\% contribution
from H$_2$ excited in PDRs.  No evidence of PAH emission is seen in the
spectra, which might have been expected based on its nebular C/O ratio.

The emission in the nebula is concentrated in small knots and clumps
throughout the main rings. There is an anticorrelation between the H$_2$
emission and the ionized gas as traced by H$\alpha$ images, indicating a
segregation of the molecular and ionized gas on the scale of the observed
clumps.  The IRAC images resolve the extensively studied cometary knots in
the inner region of the nebula.  The tails of the knots and the radial
rays extending into the outer regions of the PN are seen in emission in
the IRAC bands.  In the Northeast Arc, the H$_2$ emission is located in a
shell outside of the H$\alpha$ emission.



\acknowledgments

We gratefully thank Mark Wolfire and Michael Kaufman for providing us with
their latest calculations of the H$_2$ line strengths.  We thank M. S.
Kelley for providing his IRS extended source correction to us. This work
is based in part on the IRAC post-BCD processing software ``IRAC\_proc"
developed by Michael Schuster, Massimo Marengo and Brian Patten at the
Smithsonian Astrophysical Observatory.  This work is based in part on
observations made with the Spitzer Space Telescope, which is operated by
the Jet Propulsion Laboratory, California Institute of Technology under
NASA contract 1407. Support for this work was provided by NASA through
Contract Number 1256790 issued by JPL/Caltech. Support for the IRAC
instrument was provided by NASA through Contract Number 960541 issued by
JPL. This work made use of the Two Micron All Sky Survey (2MASS) database,
which is a joint project of the University of Massachusetts and the
Infrared Processing and Analysis Center/California Institute of
Technology, funded by the National Aeronautics and Space Administration
and the National Science Foundation. HAS acknowledges partial support from
NASA Grant NAG5-10654.



Facilities: Spitzer.

\begin{table}
\begin{center}
\caption{NGC 7293 Knots and Profile Positions\tablenotemark{a}\label{knottab}}
\begin{tabular}{crrrrr}
\\
\tableline\tableline
Knot \# & Name\tablenotemark{b} & Start R.A. & Dec. (J2000) & End R.A. & Dec. (J2000) \\
\tableline
1 & 356-216 & 22:29:35.76 & -20:52:12.9 & 22:29:35.20 & -20:52:35.5 \\
2 & 378-800 & 22:29:37.79 & -20:48:06.7 & 22:29:37.79 & -20:48:38.2 \\
3 & 352-815 & 22:29:35.40 & -20:48:18.9 & 22:29:34.40 & -20:47:54.5 \\
4 & 459-905 & 22:29:45.75 & -20:49:07.6 & 22:29:46.60 & -20:49:35.0 \\
\tableline
\end{tabular}
\tablenotetext{a}{The positions show the start and stop coordinates of the
profiles shown in Figure \ref{knots}.  The profile width is 2\arcsec.}
\tablenotetext{b}{
The knot ID in the scheme of \citet{odell97}, which is based on the
position of the bright head of the knot. The first number is the tenths of
the R.A. seconds coordinate, and the second number is the last digit of
the Declination coordinates minutes appended to the Declination seconds.
}
\end{center}
\end{table}


\clearpage

\clearpage

\clearpage



\clearpage

\clearpage







\end{document}